\documentclass[12pt]{article}
\usepackage{latexsym}                 
\usepackage[T1]{fontenc}
\usepackage{lmodern}
\usepackage[latin1]{inputenc}
\usepackage{latexsym}                 
\usepackage{color}                 
\usepackage{amssymb}
\usepackage{amsmath}
\usepackage{amsthm}
\usepackage{graphicx}
\usepackage{hyperref}
\usepackage{verbatim}
\usepackage{mathptmx}      % use Times fonts if available on your TeX system
\newcommand{\eq}{\end{equation}}
\def\dfrac#1#2{\displaystyle{\frac{#1}{#2}   }}

\def\b1{\mathbf 1}

\def\biglf{\par\bigskip\noindent}
\def\eref#1{(\ref{#1}%
)}
\def\RSref#1{\ref{#1}%
}
\def\RSlabel#1{\label{#1}%
}
\def\RScite#1{\cite{#1}%
}

\newcommand{\bql}[1]{%
\begin{equation}\label{#1}%
}
\def\filename#1{}
\begin{document}
\begin{center}
 
  {\Large\bf On COVID-19 Modelling} 
  \\~\\
  Robert Schaback\footnote{Address:\\
    Prof. Dr. R. Schaback, Institut für Numerische und Angewandte Mathematik,\\
    Universität Göttingen, 
    Lotzestraße 16-18, 37083 Göttingen\\
    schaback@math.uni-goettingen.de,
    http://num.math.uni-goettingen.de/schaback/}\\~\\
  May 10, 2020
\end{center} 
This contribution analyzes the COVID-19 outbreak
by comparably simple mathematical and numerical methods.
The final goal
is to predict the peak of the epidemic outbreak per country
with a reliable technique. This is done by an
algorithm motivated by standard SIR models and aligned with the
standard data provided by the Johns Hopkins University.
To reconstruct data for the unregistered Infected,
the algorithm uses
current values of the infection fatality rate and
a data-driven estimation of a specific form
of the recovery rate. All other ingredients are data-driven as well.
Various examples of predictions are provided for illustration.  
%****************************************************************
%****************************************************************
\section{Introduction and Overview}\RSlabel{SecIntro}
This contribution starts in section \RSref{SecClasSIRMod}
with a rather elementary reconciliation
of the standard SIR model for epidemics, featuring the central
notions like {\em Basic Reproduction Number},
{\em Herd Immunity Threshold}, and {\em Doubling Time}, together with some
critical remarks on their abuse in the media. Experts can skip
over this completely. Readers interested in the predictions
should jump right away to section \RSref{SecPutFM}.
\biglf
Section \RSref{SecAvDat} describes the Johns Hopkins data source
with its limitations and flaws, and then presents a variation
of a SIR model that can be applied directly to the data. It allows to
estimate basic pa\-ra\-me\-ters, including the Basic Reproduction Number,
but does not work for predictions of peaks of epidemics.
\biglf
To
achieve the latter goal, section \RSref{SecEtaSIRESM}
combines the data-compatible model of section \RSref{SecAvDat}
with a SIR model dealing with the unknown Susceptibles
and the unregistered Infectious. This needs two extra
pa\-ra\-me\-ters that
should be extracted from the literature. The first is the
infection fatality rate, as provided e.g. by
\RScite{verity-et-al:2020-1, streeck:2020-1}, combined with the
case fatality rate that can be deduced from the Johns Hopkins data.
This pa\-ra\-me\-ter is treated in section \RSref{SecILR}, and like
\RScite{bommer-vollmer:2020-1} it gives a detection rate for the confirmed
cases.
\biglf
Section
\RSref{SecRR} introduces a version of a recovery rate
that can be directly used in the model and
estimated from the infection fatality rate and the
observable case fatality and case death rates. However, this pa\-ra\-me\-ter
is not needed for prediction,
just for determination of the unknown variables from the known data
as long as the latter are available.
\biglf
Then section \RSref{SecPutFM} combines all of this into an
algorithm that makes predictions under the assumption that there are no further
changes to the pa\-ra\-me\-ters induced by political action.
It estimates the pa\-ra\-me\-ters of a full SIR model from the available
Johns-Hopkins data 
by the techniques of section \RSref{SecEtaSIRESM},
using two additional technical pa\-ra\-me\-ters: 
the number of days used
backwards for estimation of constants,
and the number of days in which recovery or death
must be expected, for estimation of case fatality and recovery rates.
After the data-driven estimation of these pa\-ra\-me\-ters,
the prediction uses only the case fatality rate.
All other ingredients are derived from the Johns Hopkins data.
\biglf
Results are presented in section \RSref{SecPutFM}. 
Given the large uncertainties in the Johns-Hopkins data, the
predictions are rather plausible. However, reality will have
the final word on this
prediction model.
\biglf
The paper closes with a summary and a list of open problems.
 %****************************************************************
%****************************************************************
%****************************************************************
\section{Classical SIR Modeling}\RSlabel{SecClasSIRMod}
This contains some basic notions that are useful
for modelling epidemics, and that were in use in the media during the COVID-19
outbreak.  Among other things, there will be a rigid mathematical
underpinning of what is precisely meant by
\begin{itemize}
\item {\em flattening the epidemic outbreak},
\item {\em basic reproduction number},
  \item {\em Herd Immunity Threshold}, and
  \item {\em doubling time},
\end{itemize}
pointing out certain abuses of these notions.
This will not work without calculus, but 
things were kept as simple as possible.
Readers should take the opportunity to
brush up their calculus knowledge. Experts should go over to
section \RSref{SecAvDat}.
%****************************************************************
\subsection{The Model}\RSlabel{SecSIRM} 
The simplest standard ``SIR'' model of epidemics
(e.g. \RScite{hethcote:2000-1} and easily retrievable in the wikipedia
\RScite{wikipedia:SIR})
deals with three variables
  \begin{enumerate}
 \item Susceptible ($S$),
 \item Infectious ($I$), and
 \item Removed ($R$).
  \end{enumerate}
  The Removed cannot infect anybody anymore, being either dead or immune.
  This is the viewpoint of bacteria of viruses.
  The difference between death and immunity of subjects
  is totally irrelevant for them:
  they cannot proliferate anymore in both cases.
  The SIR model cannot say anything about death rates of persons.
  \biglf
  The Susceptible are not yet infected and not immune, while the
  Infectious can infect Susceptibles. The three classes
  $S,\,I$, and $R$ are disjoint and add up
  to a fixed total population count $N=S+I+R$. All of these are ideally
  assumed to be
  smooth functions of time $t$, and satisfy the differential equations
  \bql{eqSIRsys}
  \begin{array}{rcl}
    \dot{S}&=&-\beta \dfrac{S}{N}I,\\[0.4cm]
    \dot{I}&=&+\beta \dfrac{S}{N}I-\gamma I,\\[0.4cm]
    \dot{R}&=&\gamma I.
  \end{array}
  \eq
  where the dot stands for the time derivative, and where $\beta$ and $\gamma$
  are positive pa\-ra\-me\-ters. The product
  $\frac{S}{N}I$ models the probability that an Infectious meets a
  Susceptible. Note that the Removed of the SIR model are not the Recovered of
  the Johns Hopkins data that we treat later,
  and the SIR model does not account for the Confirmed counted there.
  \biglf
  Since $\dot N= \dot{S}+ \dot{I}+\dot{R}=0$,
  the equation $N=S+I+R$ is kept valid at all times. The term
  $\beta\frac{S}{N}I$ moves Susceptibles to Infectious, while $\gamma I$
  moves Infectious to Removed. Thus $\beta$ represents an {\em infection rate}
  while the {\em removal rate} $\gamma$ accounts for either healing or
  fatality after infection, i.e. immunity. Political decisions about reducing
  contact probabilities will affect $\beta$, while $\gamma$ resembles
  the balance between the medical aggressivity of the infection
  and the quality of the health care system.
  \biglf
    As long as the Infectious $I$ are positive, the Susceptibles $S$ are
  decreasing, while the Removed $R$ are increasing. Excluding the trivial
  case of zero Infectious from now on,
  the Removed and the Susceptible will be strictly monotonic.
\biglf
  Qualitatively, the system is not really dependent on $N$, because one can multiply
  $N,\,R,\,I$, and $S$ by arbitrary factors without changing the system.
  As an aside, one can also go over to relative
  quantities with the two differential equations
    $$
  \begin{array}{rcl}
    \dfrac{d}{dt}\dfrac{I}{N}
    &=&+\beta \left(1-\dfrac{I}{N}-\dfrac{R}{N}\right)\dfrac{I}{N}-\gamma \dfrac{I}{N},\\[0.4cm]
    \dfrac{d}{dt}\dfrac{R}{N}
    &=&\gamma \dfrac{I}{N}.
\end{array}
  $$
  Figure \ref{FigSIR01}
  shows some simple examples that will be explained in some detail below.
  \begin{figure}
\begin{center}
  \includegraphics[width=10.0cm,height=8.0cm]{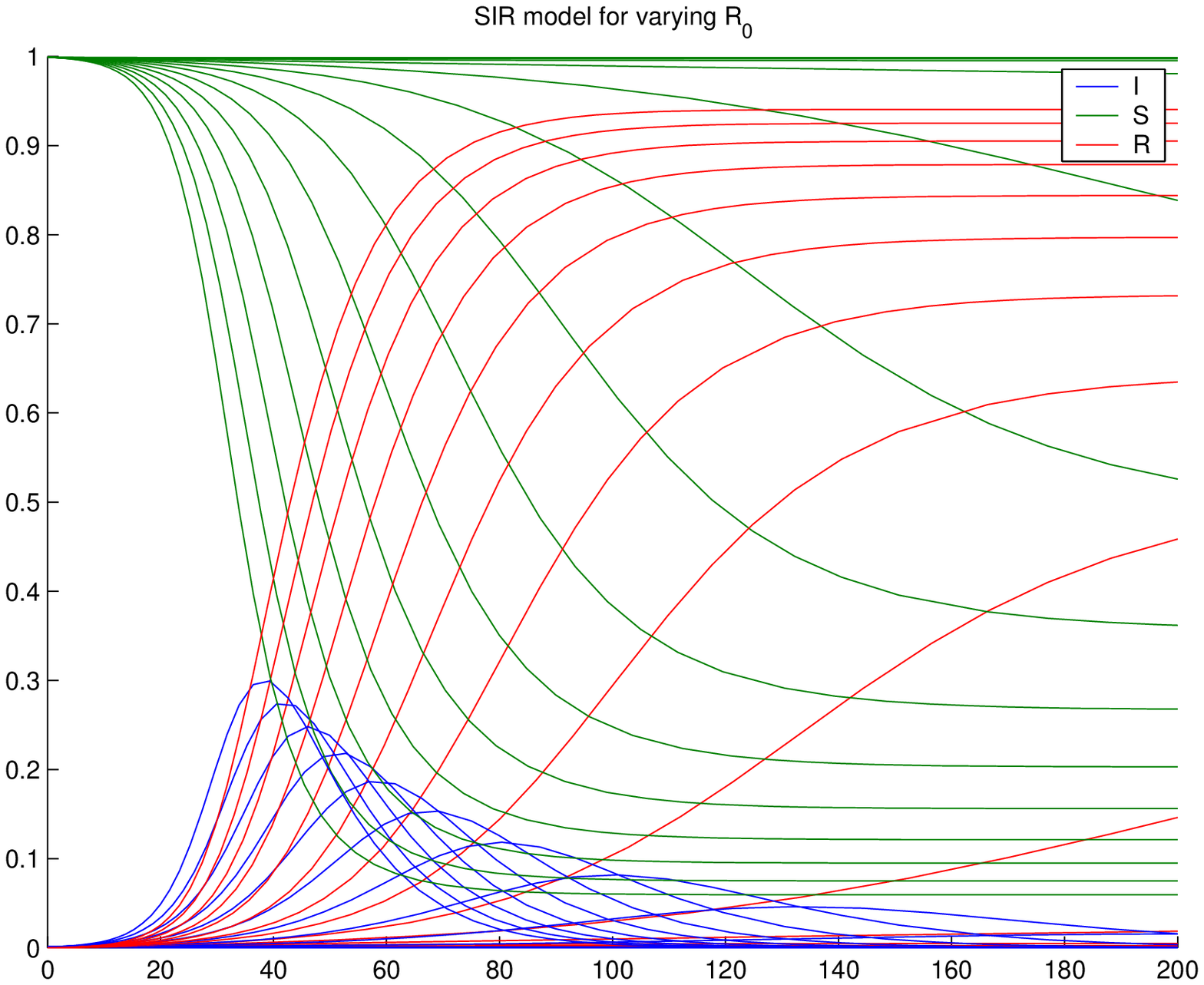}
  \caption{Some typical SIR system solutions  %\red{(Program: SIR01.m)}
    \RSlabel{FigSIR01}}
\end{center} 
\end{figure}
%****************************************************************
\subsection{Conditions for Outbreaks}\RSlabel{SecSIRIC} 

  We start by looking at the initial conditions. 
  Since everything is invariant under an additive time shift, we can start at
  time 0 and consider
  $$
 \dot{I}(0)=+\beta \dfrac{S(0)}{N}I(0)-\gamma I(0)
 $$
 and see that the Infectious decrease right from the start if
\bql{eqS0Ndecr}
\dfrac{S(0)}{N}<\dfrac{\gamma}{\beta}, 
\eq
and this keeps going on since $S$ must decrease and
\bql{eqIIbSN}
\dfrac{\dot I}{I}(t)=\beta \dfrac{S(t)}{N}-\gamma\leq \beta \dfrac{S(0)}{N}-\gamma<0.
\eq
There is no outbreak in this case, because there are not enough Susceptibles at
start time. The case $S(0)=N$ means that there is no infection at all, and we
ignore it, though it is a solution to the system, with $I=R=0,\,S=N$ throughout.
%****************************************************************
\subsection{The Peak}\RSlabel{SecPeak} 
  If there is a
  time instance $t_I$ (maybe $t_I=0$ above)
  where the Infectious are positive and
  do not change, we have
    \bql{eqImax} 
  \begin{array}{rcl}
     0=\dot{I}(t_I)&=&\beta \dfrac{S(t_I)}{N}I(t_i)-\gamma I(t_I),\\[0.4cm]
     \gamma&=&\beta \dfrac{S(t_I)}{N}\leq \beta.
  \end{array}
  \eq
  If $\beta<\gamma$ holds, this situation cannot occur, and $I$ must be
  decreasing all the time, i.e. the infection dies out. This is what everybody
  wants. There is no outbreak. In case of $\gamma=\beta$ we go back to the
  initial situation of the previous section
  and see that there is no outbreak due to $S(0)/N<1$
  if there is an infection at all. 
  \biglf
  The interesting case is $\beta >\gamma$. Then the first part of
  \eref{eqIIbSN} shows that as soon as $t$ is larger than the peak time $t_I$,
  the Infectious will decrease due to \eref{eqImax}. Therefore
  the zero of $\dot I$ must be a maximum,
  i.e. a {\em peak}, and it is unique. The Infectious go to zero
  even in the peak situation.
  \biglf
  It is one of the  most important practical problems
  in the beginning of an epidemic to predict
  \begin{itemize}
 \item
whether there will be a peak at
  all,
\item  when the possible peak will come, and
  \item how many Infectious will be at the
  peak. \end{itemize}
  This can be answered if one has good estimates
  for $\beta$ and $\gamma$, and we shall deal with this problem
  in the major part of this paper,
  \biglf
  In real life it is highly important to avoid the peak situation,
  and this can only be done by  administrative measures that change $\beta$ and
  $\gamma$ to the situation  $\beta<\gamma$.
  This is what management of epidemics
  is all about, provided that an epidemic follows the SIR model.
  We shall see how countries perform.
  %****************************************************************
\subsection{Basic Reproduction Number}\RSlabel{SecSIRBRN} 
  The quotient
  $$
R_0:=\dfrac{\beta}{\gamma}
  $$
is called the {\em basic reproduction number}. If it is not larger than one,
there is no outbreak, whatever the initial conditions are.
If it is larger than one,
there is an outbreak provided that 
\bql{eqSNgb}
1>\frac{S(0)}{N}>\frac{\gamma}{\beta}=\dfrac{1}{R_0}
\eq
holds. In that case, there is a time $t_I$
where $I$ reaches a maximum, and \eref{eqImax} holds there. When we discuss an
outbreak in what follows, we always assume $R_0>1$ and \eref{eqSNgb}. If we
later let $R_0$
tend to 1 from above, we also require  that $S(0)$ tends to $N$ from below,
in order to stay in the outbreak situation. 
\biglf
Both $\beta$ and $\gamma$  change under a change of time scale, but
the basic reproduction number is invariant. Physically, $\beta$ and $\gamma$
have the dimension $time^{-1}$, but $R_0=\beta/\gamma$ is dimensionless.
%****************************************************************
\subsection{Examples}\RSlabel{SecExa} 

Figure \RSref{FigSIR01} shows a series of
test runs with $S(0)=N\cdot 0.999$ and
$R(0)=0$
with fixed $\gamma=0.1$ and $\beta$
varying from $0.02$ to $0.3$, such that $R_0$
varies from $1/5$ to $3$. Due to the realistically small $I(0)$ being
$0.001\%$
of the population, one cannot see
the decaying cases near startup, but the tails of the blue $I$ curves
are decaying examples by starting value,
due to $\frac{S(t)}{N}<\frac{\gamma}{\beta}=1/R_0$ when
started at time $t$. Decreasing $R_0$ flattens the blue curves for $I$. One can
observe that $I$ always dies out, while $S$ and $R$ tend to fixed positive
levels. We shall prove this below. From the system, one can also infer
that $R$ has an inflection point where $I$ has its maximum, since
$\stackrel{..}{R}=\gamma \dot I$. If only $R$ would be observable, one could
locate the peak of $I$ via the inflection point of $R$.
\biglf
Figure \RSref{FigSIR02} shows an artifical case with a large starting value
$I(0)=N/2$, fixed $\gamma=0.1$ and $\beta$ varying from 0.005 to 0.3, letting
$R_0$ vary from 0.05 to 3. In contrast to Figure \RSref{FigSIR01},
this example shows cases with small $R_0$
properly. The essence is that the Infectious go down, whether
they have a peak or not, and there will always be a portion of Susceptibles.
Again, we shall prove this below.
\begin{figure}
\begin{center}
  \includegraphics[width=10.0cm,height=8.0cm]{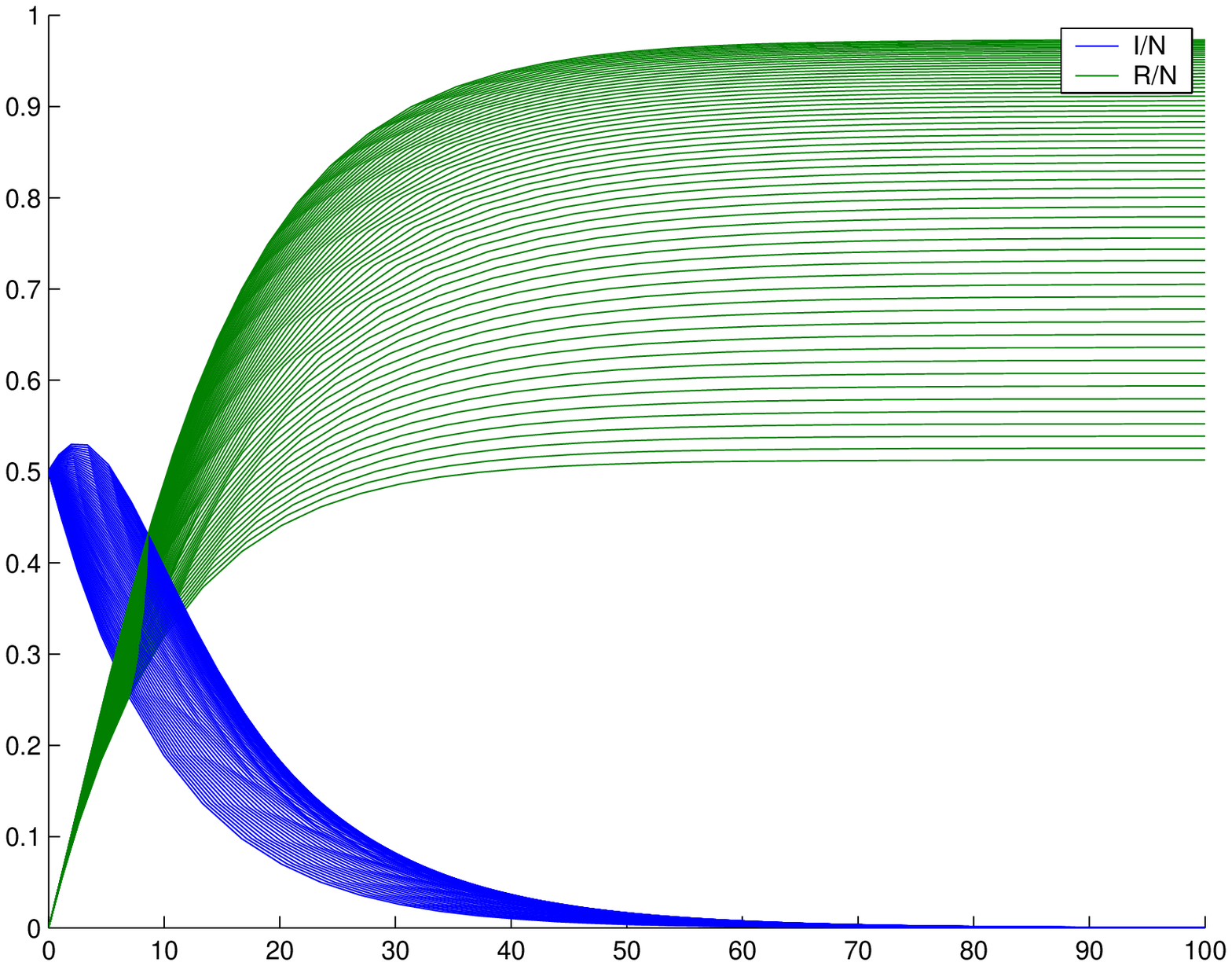}
  \caption{Some other typical SIR system solutions % \red{(Program: SIR01.m)}
    \RSlabel{FigSIR02}}
\end{center} 
  \end{figure}
%****************************************************************
\subsection{Herd Immunity Threshold}\RSlabel{SecHIT} 
This is a number related to the Basic  Reproduction Number $R_0$ by
$$
HIT=1-\dfrac{1}{R_0}
$$
following a special scenario. If a population is threatened
by an infection with Basic  Reproduction Number $R_0$, what is the
number of immune persons needed to prevent an outbreak right from the start?
We can read this off equation \eref{eqS0Ndecr}
in the ideal situation that $I(0)=0$ and $S(0)+R(0)=N$, namely
$$
\dfrac{S(0)}{N}=1-\dfrac{R(0)}{N}=\dfrac{\gamma}{\beta}=\dfrac{1}{R_0}
$$
implying
$$
\dfrac{R(0)}{N}=1-\dfrac{1}{R_0}
$$
as the threshold between outbreak and decay. This does not refer to a whole
epidemic scenario, nor to an epidemic outbreak. It is a condition
to be checked before anything happens, and useless
within a developing epidemic, whatever the media say.
\biglf
In the peak situation of \eref{eqImax}, the fraction
$$
\dfrac{R(t_I)+I(t_I)}{N}=\dfrac{N-S(t_I)}{N}=1-\dfrac{1}{R_0}
$$
of the Non-Susceptible at the peak $t_I$ of $I$
is exactly the Herd Immunity Threshold.
Thus it is correct to say that if the Immune of a population
are below the Herd Immunity Threshold at startup,
and if the Basic Reproduction Number is larger than one,
the sum of the Immune and the Infectious will rise up to
the Herd Immunity Threshold and then the Infectious will decay.
This is often stated imprecisely in the media.
\biglf
Furthermore, the Herd Immunity Threshold has
nothing to
do with the important long-term ratio of Susceptibles to Removed.
We shall address this ratio in section
\RSref{SecSIRLB}.
%****************************************************************
\subsection{Locating the Peak}\RSlabel{SecSIRLtP} 

The most interesting questions
during an outbreak with $R_0>1$ are
\begin{itemize}
\item At which time $t_I$ will we reach the maximum of the Infectious, and
\item what is $I(t_I)$, i.e. how many people will maximally be infectious at that time? 
\end{itemize}
It will turn out that there are no easy direct answers.
From \eref{eqImax} we see that at the maximum of $I$
the Susceptibles $S$ 
have the value
$$
\dfrac{\gamma}{\beta}=\dfrac{S(t_I)}{N}=\dfrac{1}{R_0},
$$
i.e. the portion $1/R_0$ of the population is susceptible. From that time on,
the Infectious decrease. In terms of $R$ and $I$, the value
$$
\dfrac{R(t_I)+I(t_I)}{N}=1-\dfrac{1}{R_0}
$$
of the Non-Susceptibles
marks the peak of the Infectious at the Herd Immunity Threshold.
``Flattening the curve'', as often mentioned
in the media, is intended to mean
making the maximum of $I$ smaller, but this is not
exactly what happens, since the maximum is described by the penultimate
equation concerning the Susceptibles,
while for $I(t_I)$ we only know
\bql{eqINbnd}
\dfrac{I(t_I)}{N}\leq\dfrac{R(t_I)+I(t_I)}{N}=1-\dfrac{1}{R_0}
\eq
yielding that the left-hand side gets smaller if $R_0$ gets closer to one.
Politically, this requires either making $\beta$ smaller via reducing contact
probabilities or making $\gamma$ larger by improving the health system,
or both. Anyway, ``flattening the curve'' works by letting $R_0$ tend
to 1 from
above, but the basic reproduction number does not directly
determine the time $t_I$ of the maximum or the value there.
We shall improve the above analysis in section \RSref{SecBttP}.
%****************************************************************
\subsection{Analyzing the Outbreak}\RSlabel{SecSIRAtO} 
In the beginning of the outbreak, $S/N$ is near to
one, and therefore
$$
   \dot{I}\approx +\beta I-\gamma I
$$
   models an exponential outbreak with exponent $\beta-\gamma>0$,
   with a solution
   $$
I(t)\approx I(t_0)\exp((\beta-\gamma)t).
   $$
If this is done in discrete time steps $\Delta t$,
one has
$$
\dfrac{I(t+\Delta t)}{I(t)}\approx\exp((\beta-\gamma)\Delta t).
$$
The severity of the outbreak is not controlled by $R_0=\beta/\gamma$, but rather
via $\beta-\gamma$. Publishing single values $I(t)$ does not give any
information about $\beta-\gamma$. 
Better is the ratio of two subsequent values 
$$
\dfrac{I(t_2)}{I(t_1)}\approx\exp((\beta-\gamma)(t_2-t_1)),
$$
and if this gets smaller over time, the outbreak gets less dramatic
because $\beta-\gamma$ gets smaller. Really useful information about
an outbreak does not consist of values and not of increments, but
of increments of increments, i.e. some second derivative information.
This is what the media rarely provided during the outbreak. 
%****************************************************************
\subsection{Doubling Time}\RSlabel{SecSIRDT} 

Another information used by media during the outbreak
is the {\em doubling time}, i.e.
how many days it takes until daily values double. This is the number $n$ in 
$$
2=\dfrac{I(t+n \Delta t)}{I(t)}
\approx \exp((\beta-\gamma)n\Delta t)=(\exp((\beta-\gamma)\Delta t)^n
$$
or
$$
n=\dfrac{\log 2}{(\beta-\gamma)\Delta t},
$$
i.e. it is inversely proportional to $\beta-\gamma$.
If political action doubles
the ``doubling time'', if halves $\beta-\gamma$. If politicians do this
repeatedly, they never reach $\beta<\gamma$, and they never escape
an exponential outbreak if they do this any finite number of times.
Extending the doubling time will never prevent a peak,
it only postpones it and hopefully flattens it.
When presenting a ``doubling time'', media should always point out that
this makes only sense during an exponential outbreak. And it is not related
to the basic reproduction number $R_0=\beta/\gamma$, but rather to the
difference $\beta-\gamma$.
%****************************************************************
\subsection{Spread of Infections}\RSlabel{SecSIRSoI} 
Media often say that the basic reproduction number $R_0$ gives the
number of persons an average Infectious infects while being infectious.
This is a rather mystical statement that needs underpinning.
The quantity
$$
\dfrac{1}{\gamma}=\dfrac{I}{\dot{R}}
$$
is a value that has the physical dimension of time.
It describes the ratio between current Infectious
and current newly Removed, and thus can be seen as the average time needed for
an Infectious to get Removed, i.e. it is the average time that an Infectious can
infect others.
Correspondingly,
$$
\dot I+\gamma I= \dot I +\dot R =\beta\dfrac{S}{N}I
$$
are the newly Infected, and therefore
$$
\dfrac{1}{\beta}\dfrac{N}{S}=\dfrac{I}{\dot I +\dot R}
$$
can be seen as the time it needs for an average Infectious to generate a new
Infectious. The ratio $\frac{\beta}{\gamma}\frac{S}{N}$ then gives how many new
Infectious can be generated by an Infectious while being infected, but this is
only close to $R_0$ if $S\approx N$, i.e. at the start of an outbreak.
A correct statement is that $R_0$ is the average number of infections
an Infectious generates while being infectious, but within an unlimited supply
of Susceptibles. 
%****************************************************************
\subsection{Long-term Behavior}\RSlabel{SecSIRLB}
Besides the peak in case $R_0>1$, it is interesting to know
the portions of the population that get either removed
(by death or immunity) or
get never in contact to the infection. This concerns
the long-term behavior of the Removed and the Susceptibles.
Figures \RSref{FigSIR01}
and \RSref{FigSIR02} demonstrate how $R$ and $S$ level out
under all circumstances shown, but is this always true,
and what is the final ratio?
\biglf
If we are at a time $t_D$ behind the possible peak at $t_I$,
or in a decay situation enforced by starting value, like in \eref{eqS0Ndecr},
we know that
$I$ must decrease exponentially to zero. This follows from 
\bql{eqIII}
(\log I)^.=\dfrac{\dot I}{I}=\beta\dfrac{S(t)}{N}-\gamma\leq \beta\dfrac{S(t_D)}{N}-\gamma<0
\eq
showing that $\log I$ must decrease linearly, or $I$ must decrease
exponentially.
Thus we get rid of the Infectious in the long run, keeping only Susceptibles
and Removed. Surprisingly, this happens independent of how large $R_0$ is.
\biglf
Dividing the first equation in \eref{eqSIRsys} by the third leads to
$$
\begin{array}{rcl}
  \dfrac{\frac{d}{dt}S}{\frac{d}{dt}R}=\dfrac{dS}{dR}=
  -\dfrac{\beta}{\gamma}\dfrac{S}{N},
\end{array}
$$
and when setting $\sigma=S/N$ and $\rho=R/N$, we get
\bql{eqsrbsg}
\dfrac{d\sigma}{d\rho}=-\dfrac{\beta}{\gamma}\sigma
\eq
with the solution
\bql{eqsigsigexp}
\sigma(\rho)=\sigma(0)\exp\left(-\dfrac{\beta}{\gamma}\rho\right)
\eq
when assuming $R(0)=0$ at startup.
Since $\rho$ is increasing, it has a
limit $0<\rho_\infty\leq 1$ for $t\to\infty$,
and in this limit
$$
\sigma(\rho_\infty)=\sigma(0)\exp\left(-\dfrac{\beta}{\gamma}\rho_\infty\right)
$$
holds, together with the condition $\rho_\infty+\sigma(\rho_\infty)=1$,
because there are no more Infectious.
The equation
$$
\sigma(0)\exp\left(-\dfrac{\beta}{\gamma}\rho_\infty\right)=1-\rho_\infty
$$
has a unique solution in $(0,1)$ dependent on $\sigma(0)<1$ and
$R_0=\beta/\gamma$. See Figure \RSref{FigtestRhoInfty} for illustration.
The right-hand side $1-\rho_\infty$ of the equation is the fixed blue line,
while the green curves are the left-hand side for varying values
of $R_0=\beta/\gamma$. The intersection points are the values $\rho_\infty$ 
of $\rho$ that solve the equation. 
Looking at both sides of the equation as functions of $\rho_\infty$, an increase
of $R_0=\beta/\gamma$ for fixed $S(0)<1$ lets the intersection point move
towards 1 on the $\rho$ axis.
\biglf
Qualitatively, we can use \eref{eqsrbsg} or \eref{eqsigsigexp}
in the form
\bql{eqRbgll}
R_0=\dfrac{\beta}{\gamma}
=-\dfrac{\log(\sigma(\rho)) -\log(\sigma(0))}{\rho}
=-\dfrac{\log(\sigma_\infty)) -\log(\sigma(0))}{\rho_\infty}
\eq
to see that the ratio of Removed to Susceptibles increases with
$R_0$, but there is a logarithm involved.
\biglf
All of this has some serious implications, if the model is correct for an
epidemic situation. First, the Infectious always go to zero, but Susceptibles
always remain. This means that a new infection can always arise
whenever some infected person enters the sanitized population.
The outbreak risk is dependent on the portion $\sigma_\infty=1-\rho_\infty$
of the Susceptibles. This illustrates the importance of vaccination, e.g. 
against measles or influenza.
\biglf
The above analysis shows that large values of
$R_0$ lead to large relative values of Removed to Susceptible
in the limit. The consequence is that systems with large $R_0$ have a dramatic
outbreak and lead to a large portion of Removed. This is good news in case
that the rate of fatalities within the Removed is low, but very bad news
otherwise.
\biglf
When politicians try to ``flatten the curve'' by bringing
$R_0$ below 1 from some time on, this will automatically decrease the
asymptotic rate of Removed and  increase the
asymptotic rate of Susceptibles in the population. This is particularly
important if
the rate of fatalities within the Removed is high, but by the previous argument
the risk of re-infection rises due to the larger portion of Susceptibles.
\biglf
The decay situation \eref{eqIII} implies that
$$
\sigma_\infty=\dfrac{S(\infty)}{N}\leq \dfrac{\gamma}{\beta}=\dfrac{1}{R_0}
$$
and consequently
$$
\rho_\infty=1-\sigma_\infty\geq 1-\dfrac{1}{R_0}=HIT.
$$
Therefore the final rate of the Removed is not smaller than
the Herd Immunity Threshold.  This is good news for possible re-infections,
but only if the death rate among the Removed is small enough.
  \begin{figure}
\begin{center}
  \includegraphics[width=10.0cm,height=8.0cm]{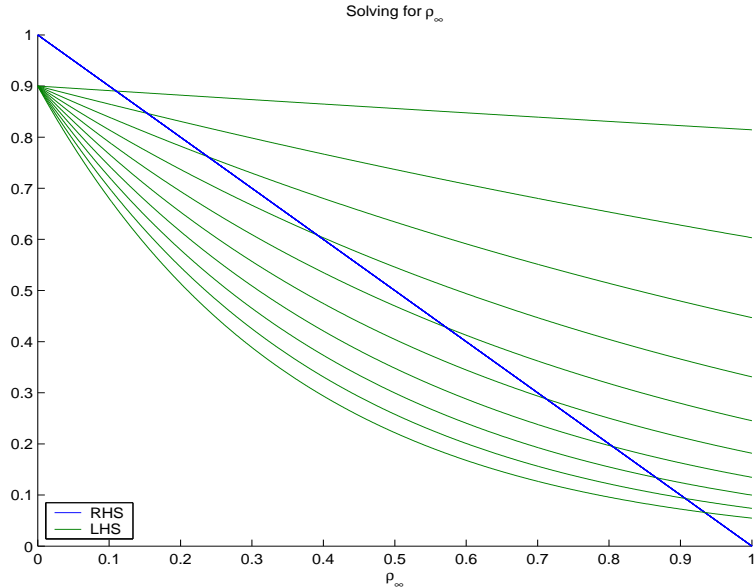}
  \caption{Solving for $\rho_\infty$ for fixed $C(0)=0.9$ and varying $R_0$
    % \red{(Program: Testrhoinfty.m)}
    \RSlabel{FigtestRhoInfty}}
\end{center} 
  \end{figure}
%****************************************************************
\subsection{Asymptotic Exponential Decay}\RSlabel{SecED}
In a decay situation like in 
\eref{eqIII}, we get
$$
I(t_D)\exp(-(\gamma-\beta\sigma_\infty)(t-t_D))\leq I(t)
\leq I(t_D)\exp(-(\gamma-\beta\sigma(t_D))(t-t_D))
$$
 to see that the exponential decay is not ruled by $\beta-\gamma$ as in
 the outbreak case with $R_0>1$, but rather by
 $-\gamma+\beta\sigma_\infty$. This also holds for large $R_0=\beta /\gamma$
 because $\sigma_\infty$ counteracts. The bell shapes
 of the peaked $I$ curves are not symmetric with respect to the peak. 
%****************************************************************
\subsection{Back to the Peak}\RSlabel{SecBttP}
  If we go back to analyzing the peak of $I$ at $t_I$ for $R_0>1$, we know
  $$
  \sigma(t_I)=\dfrac{\gamma}{\beta}=\dfrac{1}{R_0}=\sigma(\rho(t_I))
  =\sigma(0)\exp\left(-R_0\rho(t_I)\right)
  $$
  and get
  $$
\rho(t_I)=\dfrac{1}{R_0}\log(\sigma(0)R_0)
$$
leading to
$$
\dfrac{I(t_I)}{N}=1-\sigma(t_I)-\rho(t_I)=1-\dfrac{1}{R_0}-\dfrac{1}{R_0}\log(\sigma(0)R_0)
$$
as the exact value at the maximum, improving \eref{eqINbnd}.
Note that the final log is
positive due to the condition \eref{eqSNgb} for an outbreak.
\biglf
For standard infections that have starting values $\sigma(0)=S(0)/N$
very close to one,
the maximal ratio of Infectious is
$$
\dfrac{I(t_I)}{N}\approx 1-\dfrac{1}{R_0}-\dfrac{1}{R_0}\log(R_0).
$$
Figure \RSref{FigImaxR} shows the behaviour of the function,
and this is what ``flattening the curve'' is all about. A value
of $R_0=4$ gets a maximum of more than 40\% of the population infectious
at a single time. If 5\% need hospital care,
this implies that a country needs hospital beds for 2\%
of the population. The dotted line leaves the log term out, i.e.
is marks the rate of the Susceptibles at the peak,
and by \eref{eqINbnd} the difference is the rate $R(t_I)/N$ of the Recovered
at the peak.
  \begin{figure}
\begin{center}
  \includegraphics[width=10.0cm,height=8.0cm]{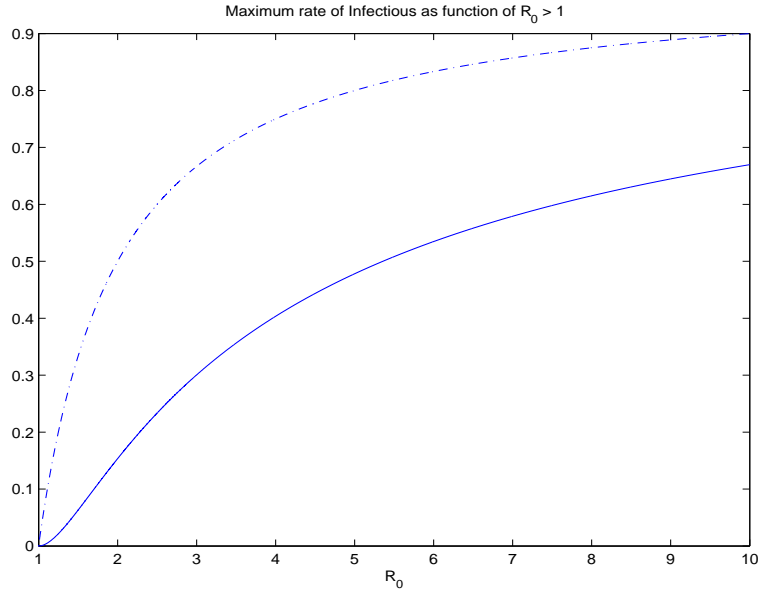}
  \caption{The effect of $R_0$ on the maximum rate of Infectious
    within the population % \red{(Program: Imax.m)}
    \RSlabel{FigImaxR}}
\end{center} 
  \end{figure}
\biglf
To analyze the peak time $t_I$, we use
$$
\dfrac{\dot I}{I}=\beta \dfrac{S}{N}-\gamma\leq\beta-\gamma 
$$
to get an upper bound for the exponential outbreak
$$
I(t)\leq I(0)\exp((\beta-\gamma)t)
$$
that implies a lower bound for $t_I$ of the form
$$1-\dfrac{1}{R_0}-\dfrac{1}{R_0}\log(R_0)\leq
1-\dfrac{1}{R_0}-\dfrac{1}{R_0}\log(\sigma(0)R_0)=
\dfrac{I(t_I)}{N}\leq \dfrac{I(0)}{N}\exp((\beta-\gamma)t_I).
$$
This needs improvement. 
% \red{To be improved ???}
%****************************************************************
%****************************************************************
\subsection{Flattening the Curve}\RSlabel{SecBtFtC}
 To get a quantitative result about ``flattening the curve'', we first evaluate the integral
$$
\int_0^\infty \dfrac{I(s)}{N}ds=\dfrac{1}{\gamma}\int_0^\infty \dfrac{R'(s)}{N}ds=\dfrac{1}{\gamma}\rho(\infty)
$$
assuming $R(0)=0$,
and set it equal to an integral over the constant value at the maximum, i.e. we
squeeze the area under the curve into a rectangle of length $b-a$ under the maximal value, i.e.
$$
\dfrac{1}{\gamma}\rho(\infty)=(b-a)\dfrac{I(t_I)}{N}.
$$
This implies
$$
b-a=\dfrac{\dfrac{1}{\gamma}\rho(\infty)}{\dfrac{I(t_I)}{N}}
\geq \dfrac{1}{\gamma}\dfrac{\rho(\infty)}{1-\frac{1}{R_0}},
$$
and if we ``flatten the curve'' by letting $R_0$ tend to 1 from above,
we see that the length $b-a$ of the above rectangle goes to infinity like
$R_0/(1-R_0)$, because $\rho_\infty$ tends to 1.
\biglf
If there is no peak, e.g. if $R_0=\beta/\gamma$ is below 1 either at the
beginning or after some political intervention, one can repeat the above
argument starting with the Infectious at some time $t$ looking at the
area under $I$ from $t$ to infinity:
$$
\int_t^\infty \dfrac{I(s)}{N}ds=\dfrac{1}{\gamma}\int_t^\infty \dfrac{R'(s)}{N}ds=\dfrac{1}{\gamma}(\rho(\infty)-\rho(t))=(b-a)\dfrac{I(t)}{N}
$$
leading to
$$
b-a=\dfrac{1}{\gamma}\dfrac{\rho(\infty)-\rho(t)}{\dfrac{I(t)}{N}}
= \dfrac{1}{\gamma}\dfrac{\rho(\infty)-\rho(t)}{1-\sigma(t)-\rho(t)}
$$
%\red{For $t\to\infty$, this is 0/0 and needs some
%L'Hospital argument. To be done ...}
This needs improvement as well.
%****************************************************************
\subsection{The Infection Timescale}\RSlabel{SecIT}
Here is a detour that is well-known in the SIR literature.
The SIR system can be written as
$$
\begin{array}{rcl}
  \dfrac{dS}{N}&=& -\beta \dfrac{S}{N} \dfrac{I}{N}dt\\[0.4cm]
  \dfrac{dI}{N}&=& \left(+\beta \dfrac{S}{N}-\gamma\right)
  \dfrac{I}{N}dt\\[0.4cm]
  \dfrac{dR}{N}&=&\gamma \dfrac{I}{N}dt
\end{array}
$$
and in a new time variable $\tau$ with $d\tau=\frac{I}{N}dt$,
one gets the system
$$
\begin{array}{rcl}
  \dfrac{d\sigma}{d\tau}&=& -\beta \sigma(\tau),\\[0.4cm]
  \dfrac{d\rho}{d\tau}&=& -\gamma
\end{array}
$$
for $\sigma=S/N$ and $\rho=R/N$ as functions of the new
{\em infection timescale } $\tau$ that one can fix as
$$
\tau(t)=\int_0^t\dfrac{I(s)}{N}ds
$$
to make sure that $\tau(0)=0$.
This implies
$$
\begin{array}{rcl}
  \sigma(\tau)&=& \sigma(0)\exp(-\beta \tau),\\
  \rho(\tau) &=& \rho(0)+\gamma\tau.
\end{array} 
$$
The beauty of this is that the roles
of $\beta$ and $\gamma$ are perfectly split, like the roles
of $\sigma$ and $\rho$.
In the new timescale, $\rho$ increases linearly and $\sigma$ decreases
exponentially. The Basic Reproduction Number then describes the fixed ratio
$$
\dfrac{\beta}{\gamma}=R_0=\dfrac{\log S(\tau)-\log S(0)}{\rho(\tau)-\rho(0)},
$$
like in \eref{eqRbgll},
and the result \eref{eqsigsigexp}  of section \RSref{SecSIRLB} comes back as
$$
\sigma(\tau)=\sigma(0)\exp\left(-\dfrac{\beta}{\gamma}\rho(\tau)  \right)
$$
for the case $\rho(0)=0$. This approach has the disadvantage to conceal the peak
within the new timescale, and is useless for peak prediction.
%****************************************************************
\subsection{Estimating and Varying Parameters}\RSlabel{SecEVP}
If data for the SIR model were
fully available, one could solve for
\bql{eqgRIbbR}
\begin{array}{rcl}
  \gamma&=&\dfrac{\dot R}{I}\\[0.4cm]
  b:=\beta\dfrac{S}{N}
  &=& \dfrac{\dot I+\gamma I}{I}=\dfrac{\dot I+\dot R}{I},\\[0.4cm]
  \beta&=&\dfrac{N}{N-I-R}\cdot\dfrac{\dot I+\dot R}{I},\\[0.4cm]
  R_0&=&\dfrac{N}{N-I-R}\cdot\dfrac{\dot I+\dot R}{\dot R}=
  -\dfrac{N}{S}\dfrac{\dot S}{\dot R}
  =-\dfrac{1}{\sigma}\dfrac{d\sigma}{d\rho},
\end{array} 
\eq
and we shall use this in section \RSref{SecJHDM}. The validity of a SIR
model can be tested by checking whether the right-hand sides
for $\beta,\;\gamma$ and $R_0$ are roughly constant.
If data are sampled locally, e.g. before or after a peak,
the above technique should determine the pa\-ra\-me\-ters for the
global epidemic, i.e. be useful for either prediction
or backward testing.
\biglf
However, in pandemics like
COVID-19, the pa\-ra\-me\-ters $\beta$ and $\gamma$ change over time
by administrative action. This means that they should be considered
as functions in the above equations, and then their changes may
be used for conclusions about the influence of such actions.
This is intended when media say that ``$R_0$ has changed''.
From this viewpoint, one can go back to the SIR model
and consider $\beta$ and $\gamma$ as ``control functions''
that just describe the relation between the variables.
\biglf
But the main argument against using \eref{eqgRIbbR} is that
the data are hardly available. This is the concern of the next section.
%********************************************************
%%********************************************************
%********************************************************
%********************************************************
%********************************************************
%********************************************************
\section{Using Available Data}\RSlabel{SecAvDat}
Now we want to confront the modelling of the previous section with available
data. This is crucial for maneuvering countries through the epidemics
\RScite{ZEIT:2020-04-16}\footnote{Original text in German, April 16$^{th}$:
  {\it Schnelle Modelle, die dem Abgleich mit der
    Wirklichkeit standhalten, sind eine
wichtige Voraussetzung, das Land politisch durch die Seuche zu steuern.}
}.
%********************************************************
\subsection{Johns Hopkins Data}\RSlabel{SecJHDat}
In this text, we work with the COVID-19 data 
from the  CSSEGISandData  repository of the Johns Hopkins University
\RScite{johns-hopkins:2020-1}. They are the only source that provides
comparable data on a worldwide scale. 
\biglf
 The numbers there are
 \begin{enumerate}
 \item Confirmed ($C$) or {\em cumulative infected} 
 \item Dead ($D$), and
 \item Recovered ($R$)
 \end{enumerate}
 as cumulative
 integer valued time series in days from Jan. 22nd, 2020. All these values
 are absolute numbers, not relative to a total population.
 Note that the unconfirmed cases are not accessible at all, while
 the Confirmed contain the Dead and the Recovered of earlier days.
 \biglf
 At this point, we do not question the integrity of the data,
 but there are many well-known flaws.
 In particular, the values
 for specific days are partly belonging to previous days,
 due to delays in the chains of data transmission in different countries.
 This is why, at some points, we shall apply some conservative smoothing of the
 data. Finally, there are inconsistencies that possibly need data changes.
 For an example, consider that usually COVID-19 cases lead to recovery or death
 within a rather fixed period, e.g. $k\approx 15-21$ days.  But some Johns Hopkins data
 have less new Infectioned at day $n$ than the sum of Recovered and Dead at day $n+k$. 
 And, there are countries like Germany who deliver data of Recovered
 in a very questionable way. The law in Germany did
 not enforce authorities to collect data
 of Recovered, and the United Kingdom did not report numbers of Dead and
 Recovered from places outside the National Health System, e.g. from Senior's
 retirement homes. Both strategies have changed somewhat in the meantime,
 as of early May, but the data still keep these flaws.
 \biglf
 We might assume that the Dead plus the  Recovered of the Johns Hopkins data
 are the Removed of the SIR model, and that the Infectious $I=C-R-D$
 of the Johns Hopkins data are the Infectious of the SIR model.
 But this is not strictly
 valid, because registration or confirmation come in the way. 
 \biglf
 On the other hand, one can take the radical viewpoint that facts are not
 interesting if they do not show up in the Johns Hopkins data.
 Except for the United Kingdom,
 the important
 figures concern COVID-19 casualties that are actually registered as such,
 others do not count,
 and serious cases needing hospitalization or leading to death
 should not go unregistered. If they do in certain countries,
 using such data will not be of any help,
 unless other data sources are available. If SIR modelling does not work
 for the Johns Hopkins data, it is time to modify the SIR technique
 appropriately, and this will be tried in this section.
 \biglf
 An important point for what follows is that  the data come as daily values.
 To make this compatible with differential equations, we shall
 replace derivatives by differences.
%****************************************************************
\subsection{Examples}\RSlabel{SecJHExa}
To get a first impression about the Johns Hopkins data,
Figure \RSref{FigJH109} shows raw
data up to day 109 (May 10$^{th}$, as of this writing).
The presentation is logarithmic, because then linearly increasing or decreasing
parts correspond to exponentially increasing or decreasing numbers in the real
data. Many presentations in the media are non-logarithmic, and then all
exponential outbreaks look similar. The real interesting data are the
Infectious $I=C-R-D$ in black that show a peak or not. The other curves
are cumulative. The data for other countries tell similar stories
and are suppressed.
\biglf
The media, in particular German TV, present COVID-19 data
in a rather debatable way. When mentioning Johns Hopkins data,
they provide $C$, $D$, and $R$ separately without stating
the most important figures, namely $I=C-D-R$, their change, and the change
of their change. When mentioning data of the Infectious
from the Robert Koch institute
alongside, they do not say precisely that
these are non-cumulative and should be compared to the $I=C-R-D$ data
of the Johns Hopkins University. And, in most cases
during the outbreak, they did not mention the change of the change.
Quite like all other media. 
\biglf
One can see in Figure \RSref{FigJH109}
that Germany and South Korea have passed the peak of the Infectious,
while France is roughly at the peak and the United States are still
in an exponential outbreak. The early figures, below day 40, are rather useless,
but then an exponential outbreak is visible in all cases.
This outbreak changes its slope
due to political actions, and we shall analyze this later.
See \RScite{dehning-et-al:2020-1} for a detailed early analysis
of slope changes.
\biglf
There are strange anomalies in the Recovered (green). France seems not to have
delivered any data between days 40 and 58, Germany changed the data delivery
policy between days 62 and 63, and the UK data for the Recovered are a
mess.
\biglf
It should be noted that
the available medical
results on the COVID-19 disease often state that
Confirmed will die or survive after a more or less fixed number of days,
roughly 14 to 21. This would imply that the red curves for the Dead and
the green curves
for the Recovered should roughly follow the blue curves for the Confirmed
with a fixed but measurable delay. This is partially observable, but
much less accurately for the Recovered. 
\begin{figure}
\begin{center}
  \includegraphics[width=10.0cm,height=5.0cm]{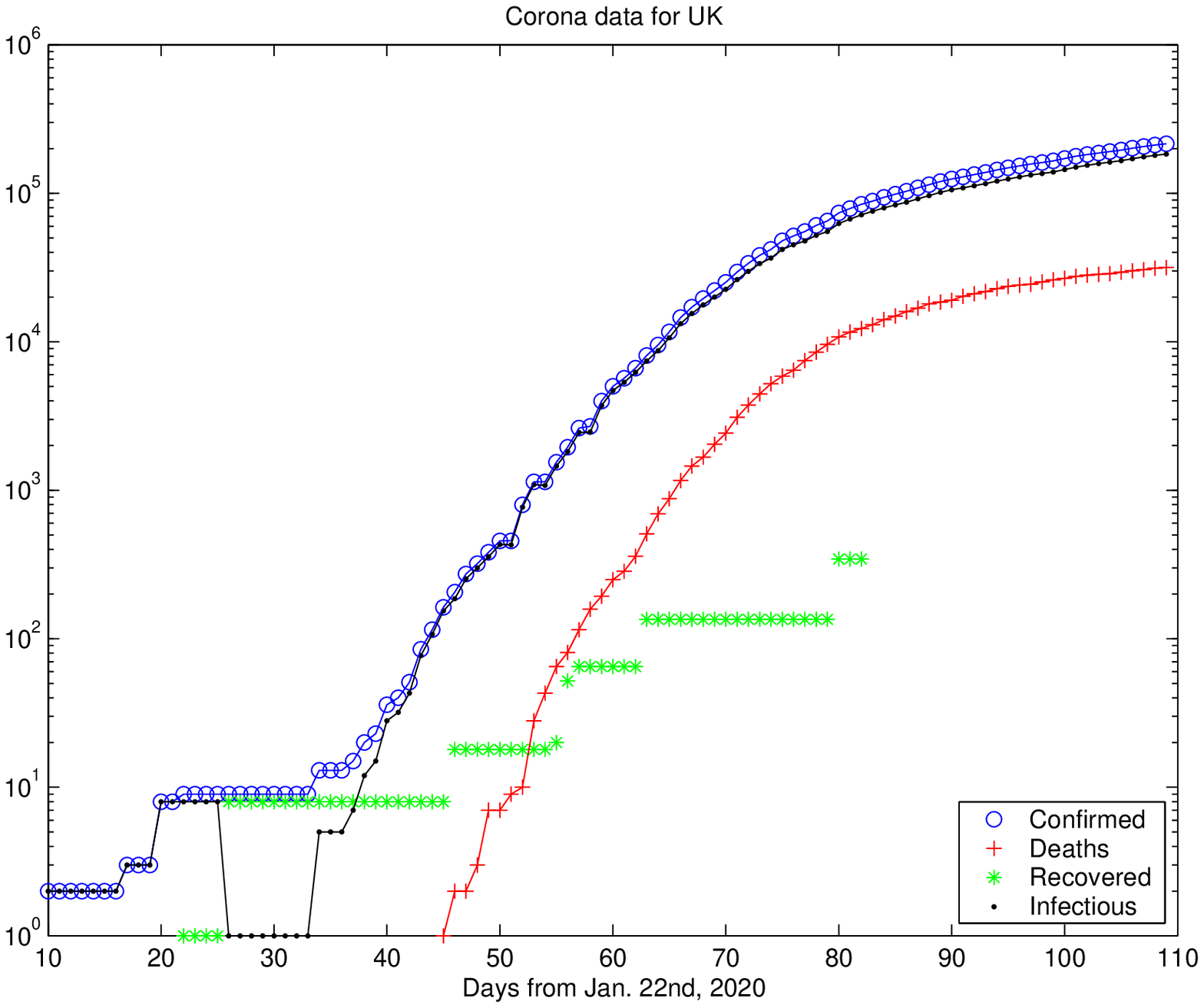}\\
  \includegraphics[width=10.0cm,height=5.0cm]{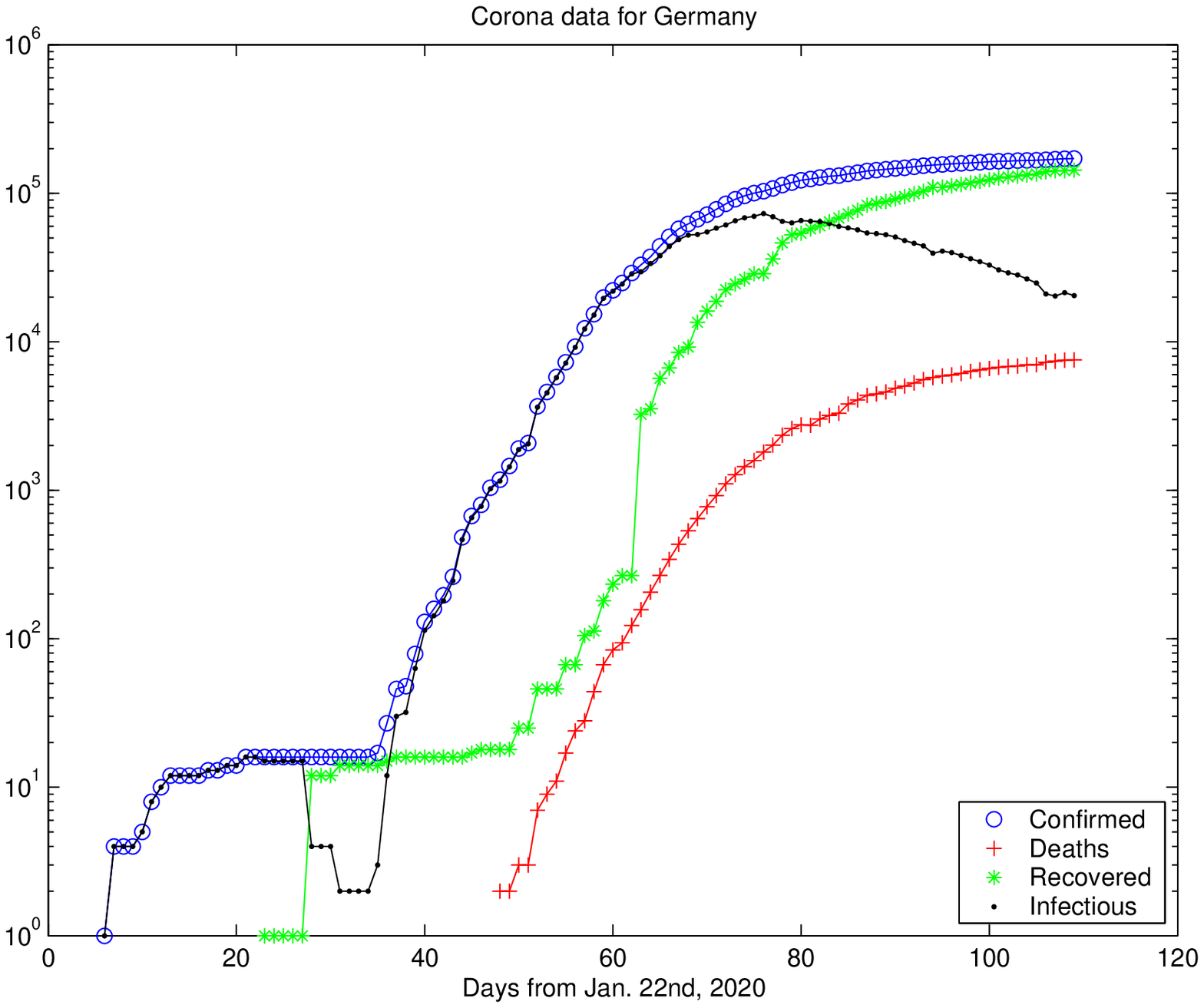}\\
  \includegraphics[width=10.0cm,height=5.0cm]{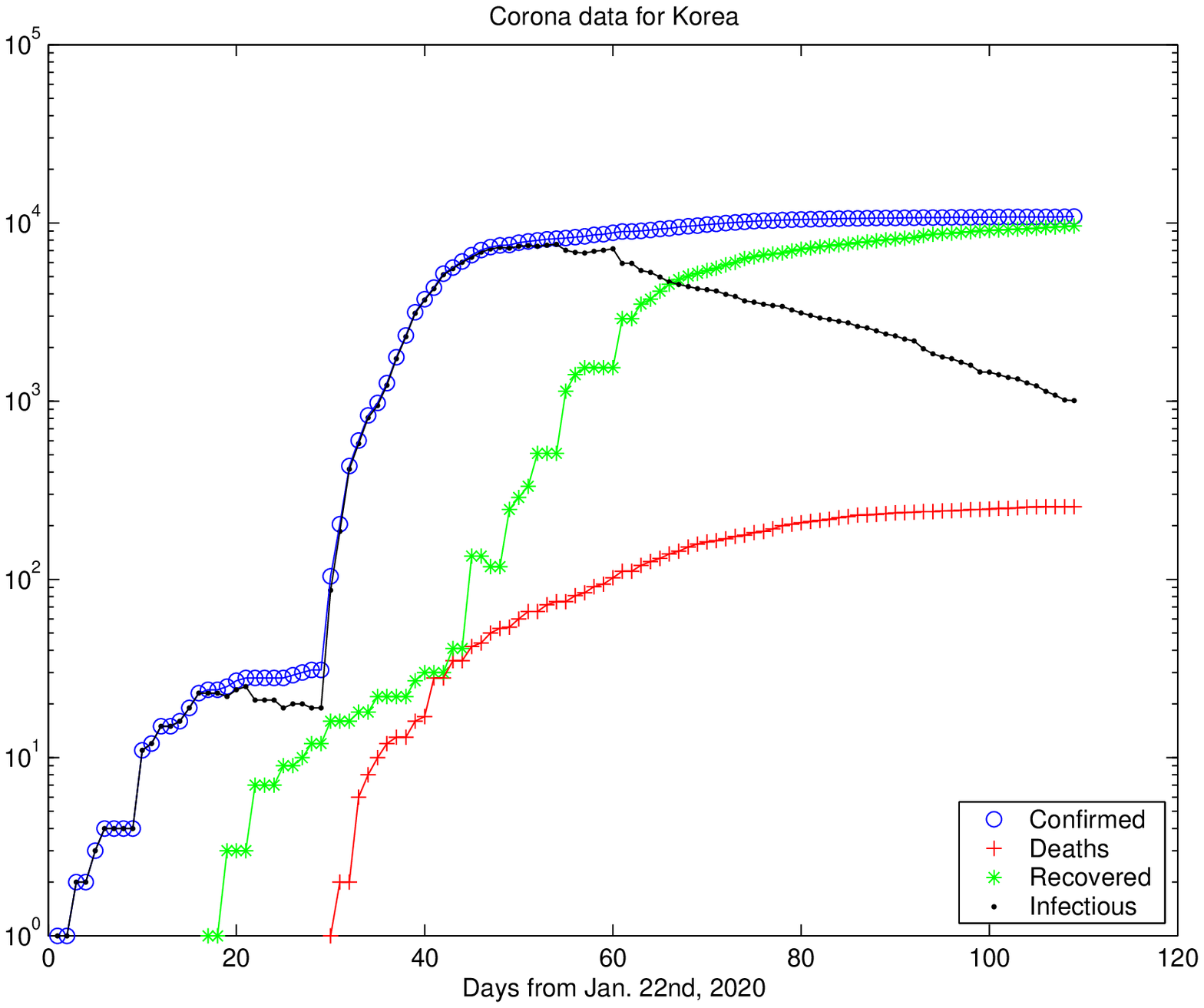}
  \includegraphics[width=10.0cm,height=5.0cm]{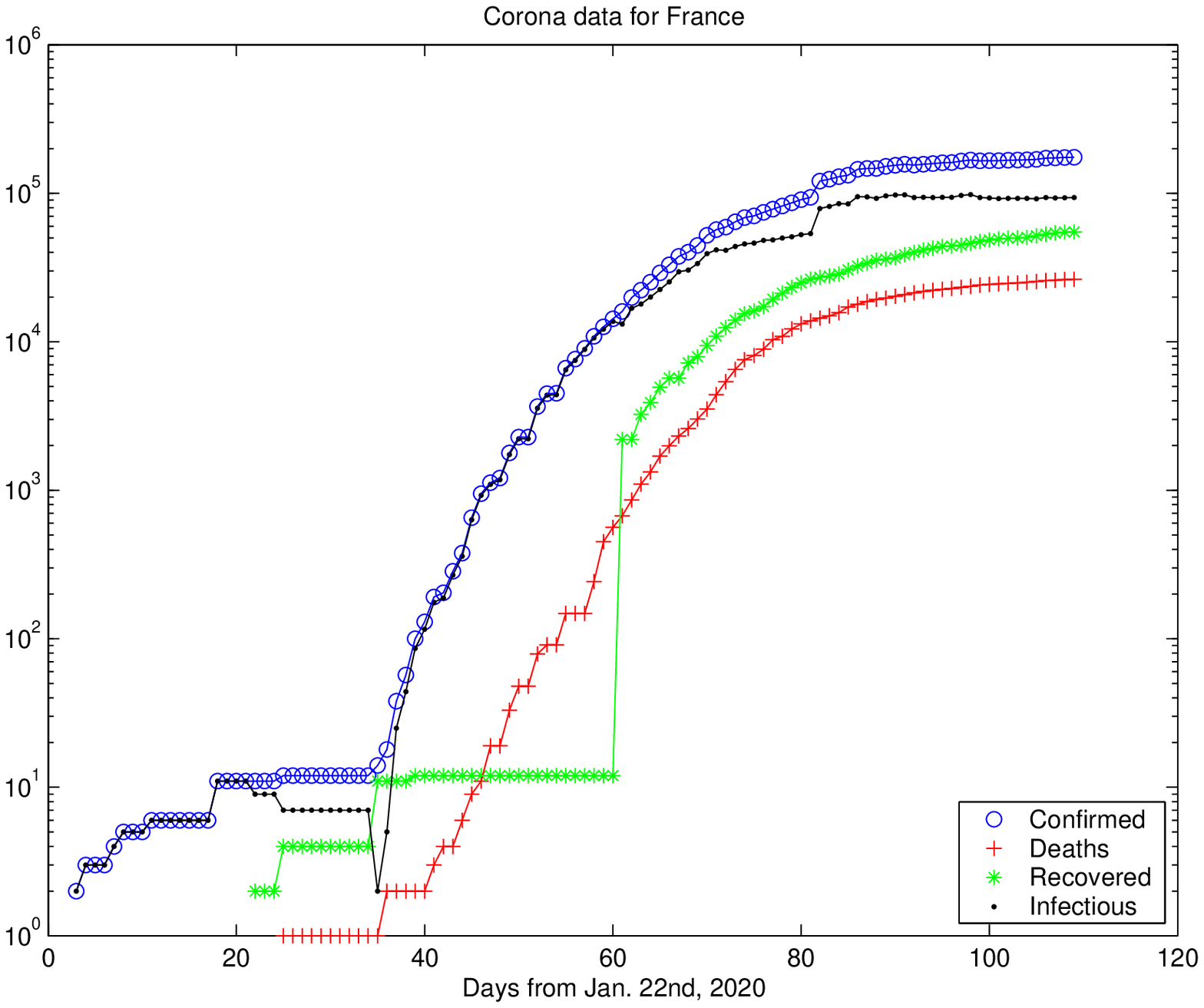}
  \caption{Raw Johns Hopkins data in logarithmic presentation up to day 109,
    from top: UK, Germany, South Korea, and France
    %\red{(Program: BasicCRDISingle.m)}
    \RSlabel{FigJH109}}
\end{center} 
  \end{figure}

%****************************************************************
\subsection{The Johns Hopkins Data Model}\RSlabel{SecJHDM}
The idea is to define a model that works exclusively
with the Johns Hopkins data, but comes close to a SIR model, without being
able to use $S$. Since the SIR model does not distinguish
between recoveries and deaths, we set
$$
R_{SIR}\Leftrightarrow D_{JH}+R_{JH}
$$
and let the Infectious be comparable, i.e.
$$
I_{SIR} \Leftrightarrow  I_{JH}:=C_{JH}-D_{JH}-R_{JH}
$$
which implies
$$
(I+R)_{SIR}\Leftrightarrow C_{JH},
$$
and we completely omit the Susceptibles. 
From now on,
we shall omit the subscript $JH$ when we use the Johns Hopkins data,
but we shall use $SIR$ when we go back to the SIR model.
\biglf
Now we go back to \eref{eqgRIbbR}
in section \RSref{SecEVP} and use
$$
\begin{array}{rcl}
  \gamma&=&\dfrac{\dot R_{SIR}}{I_{SIR}}\\[0.4cm]
  &\approx & \dfrac{(D+R)_{n+1}-(D+R)_n}{I_n}=:\gamma_n\\[0.4cm]
  b:=\beta\dfrac{S_{SIR}}{N}
  &=& \dfrac{\dot I_{SIR}+\gamma I_{SIR}}{I_{SIR}}
  =\dfrac{\dot I_{SIR}+\dot R_{SIR}}{I_{SIR}},\\[0.4cm]
  &\approx&
  \dfrac{C_{n+1}-C_n}{I_n}=:b_n,
\end{array} 
$$ 
defining time series $\gamma_n$ and $b_n$ that model $\gamma$
and $b=\beta\cdot S_{SIR}/N$ without knowing $S_{SIR}$.
This is equivalent to the
model
$$
\begin{array}{rcl}
  C_{n+1}-C_n &=& b_nI_n,\\
  I_{n+1}-I_n&=&b_nI_n-\gamma_n I_n,\\
  (R+D)_{n+1}-(R+D)_n-&=& \gamma_n I_n\\
\end{array} 
$$
that maintains $C=I+R+D$, and we may call it
a {\em Johns Hopkins Data Model}.
It is very close to a SIR model if the time
series $b_n$ is not considered to be constant, but just an approximation of
$\beta\cdot S_{SIR}/N$.
%****************************************************************
\subsubsection{Estimating $R$}\RSlabel{SecEstR}
By brute force, one can consider
\bql{eqrbg}
r_n=\dfrac{b_n}{\gamma_n}=\dfrac{C_{n+1}-C_n}{R_{n+1}+D_{n+1}-R_n-D_n}
\eq
as a data-driven substitute for
$$
\frac{\beta}{\gamma} \frac{S_{SIR}}{N}=R_0\frac{S_{SIR}}{N}.
$$
Then there is a rather simple observation:
\begin{itemize}
 \item[] If $r_n$ is smaller than one, the Infectious decrease.
\end{itemize}
It follows from
$$
\begin{array}{rcl}
  C_{n+1}-C_n&=&r_n(R_{n+1}+D_{n+1}-R_n-D_n)\\
  I_{n+1}-I_n+R_{n+1}-R_n+D_{n+1}-D_n&=&r_n(R_{n+1}+D_{n+1}-R_n-D_n)\\
%   I_{n+1}-I_n&\leq& r(R_{n+1}+D_{n+1}-R_n-D_n)-R_{n+1}+R_n-D_{n+1}+D_n\\
  I_{n+1}-I_n&=& -(1-r_n)(R_{n+1}-R_n+D_{n+1}-D_n),
\end{array}
$$
but this is visible in the data anyway
and not of much help.
\biglf
Since $r_n$ models $R_0\frac{S_{SIR}}{N}$, it always underestimates $R_0$.
This underestimation gets dramatic when it must be assumed that $S_{SIR}$
gets seriously smaller than $N$. 
\biglf
At this point, it is not intended
to forecast the epidemics. The focus is on extracting
pa\-ra\-me\-ters from the Johns Hopkins data that relate to a background SIR-type
model. 
%****************************************************************
\subsubsection{Example}\RSlabel{SecER0Ex}
Figure \RSref{FigR109} shows $R_0\frac{S_{SIR}}{N}$
estimates via $r_n$ for the last four
weeks before day 109, i.e. March 10$^{th}$.
Except for the United States, all countries
were more or less successful in pressing $R_0$ below one.
In all cases, $S_{SIR}/N$ is too close to one to
have any influence. The variation in $r_n$ is not due to the 
decrease in $S_{SIR}/N$, but should rather be attributed to political action.
As mentioned above, the estimates for $R_0$ by $r_n$
are always too optimistic. 
\biglf
For the figure, the raw Johns Hopkins data were
smoothed by a double action of
a $1/4, 1/2, 1/4$ filter on the logarithms of the data.
This smoother keeps constants
and linear sections of the logarithm invariant, i.e. it does not change
local exponential behavior. This smoothing was not applied to Figure
\RSref{FigJH109}. It was by far not strong enough to
eliminate the apparent 7-day oscillations that
are frequently in the Johns Hopkins data, see the figure.
Data from the Robert Koch Institute in Germany have even stronger
7-day variations. 
  \begin{figure}
\begin{center}
  \includegraphics[width=14.0cm,height=10.0cm]{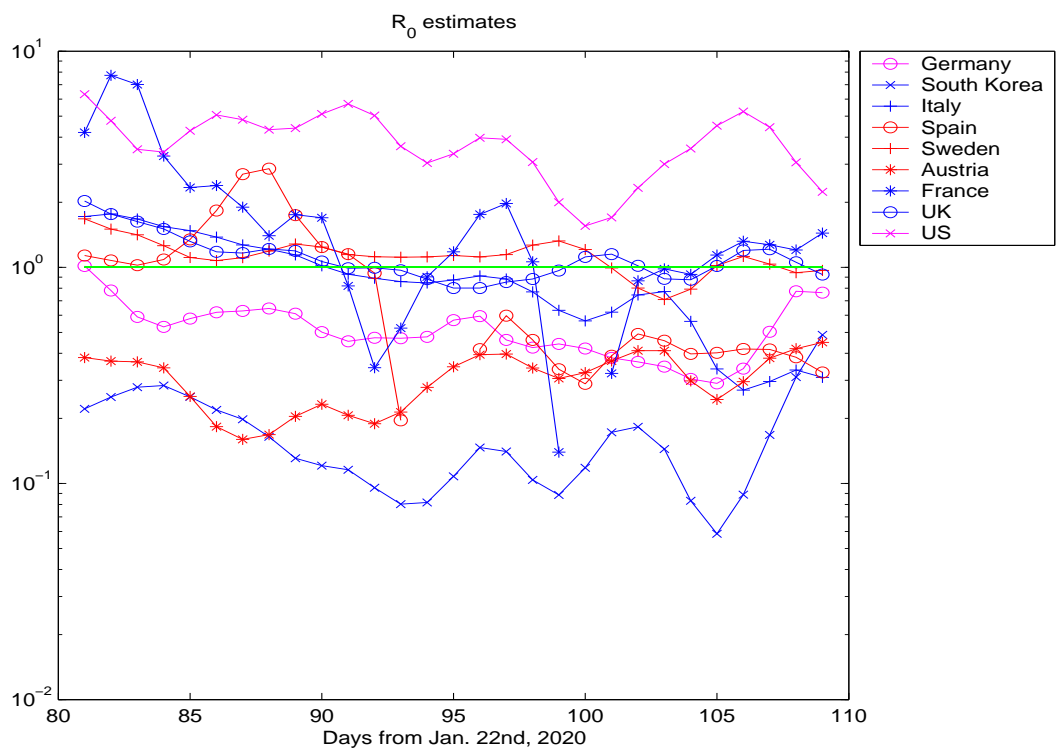}
  \caption{Estimates of $R_0$ via the time series $r_n$
    % \red{(Program: BasicCRDI\_R.m)}
    \RSlabel{FigR109}}
\end{center} 
  \end{figure}
%****************************************************************
\subsubsection{Properties of the Model}\RSlabel{SecPotM}
As long as $r_n$ is roughly constant, the above approach  will
always model an exponential outbreak or decay, but never a peak,
because the difference equations are linear. It can only help the user to tell
if there is a peak ahead or behind, depending on $r_n\approx R_0$ being larger
or smaller than 1.
If $r_n$ is kept below one, the
Confirmed Infectious will not increase, causing no new threats to
the health system. Then the $S/N$ factor will not decrease substantially, and
a full SIR model is not necessary.
\biglf
On the other hand, if a country manages to keep $r_n$ smaller than some
$r=\frac{b}{\gamma}<1$, it is clear that it takes
$$
j=\dfrac{\log(I_n)}{-\log(1+b-\gamma)}
$$
steps to bring the Confirmed Infectious down to 1, if we assume
$b$ and $\gamma$ to be constant.
Making $R_0\approx \frac{b}{\gamma}<1$ small is not the best strategy.
Instead, one should maximize $\gamma-b$. 
\biglf
As long as countries keep the  $r_n$ clearly below one, e.g. below 1/2,
this would mean that $R_0\approx r_n\frac{N}{S_{SIR}}$ stays below one
if $S_{SIR}\geq N/2$, i.e. as long as the
majority of the population has not been
in contact with the SARS-CoV-2 virus. This is good news.
But observing a small $r_n$ can conceal
a situation with a large $R_0$ if $S_{SIR}/N$ is small. This is one reason
why countries need to get a grip on the Susceptibles nationwide.
\biglf
It is tempting to use the above technique for prediction
in such a way that the $b_n$ and the $\gamma_n$ are
fitted to a constant or a linear function, and using the
values of the fit for running the system into the future.
This is very close to extending the logarithmic
plots of Figure \RSref{FigJH109} by lines, using pencil and ruler,
and thus hardly interesting. 
\biglf
So far, the above argument
cannot replace a SIR model. It only interprets the available data.
However, monitoring the Johns Hopkins data in the above way
will be very useful when it comes to evaluate the effectivity of
certain measures taken by politicians. It will be highly interesting to see how
the data of Figure \RSref{FigR109} continue, in particular when
countries relax their contact restrictions.
%****************************************************************
%****************************************************************
\subsection{Extension Towards a SIR Model}\RSlabel{SecEtaSIRM}
For cases where one still has to expect $R_0>1$, e.g. UK, US and Sweden
around day 90 (see Figure \RSref{FigR109}), the challenge remains to
predict a possible peak. Using the estimates from the previous section
is questionable, because they concern the subpopulation of Confirmed
and are systematically underestimating $R_0$.
The ``real'' SIR model will have different pa\-ra\-me\-ters, and it needs
the Susceptibles to model a peak or to make the $r_n$
estimates realistic. So far, the Johns Hopkins data work
in a range where $S/N$ is still close to one, and the Susceptibles are
considered to be abundant. But the bad news for countries with $r_n>1$
is that $r_n$ underestimates $R_0$. 
\biglf
Anyway, if one trusts the above time series as approximations to $\beta$ and
$\gamma$,
one can run a SIR model, provided one
is in the case $R_0>1$ and has reasonable starting values.
But these are a problem, because the unconfirmed Infectious
% ($M$, mysterious)
and the unconfirmed Recovered %
% ($H$, healed)
are not known, even if, for
simplicity, one assumes that there are no unconfirmed COVID-19 deaths.
\biglf
For an unrealistic scenario, consider {\em Total Registration},
i.e. all Infected are automatically confirmed.
Then the Susceptibles
in the Johns Hopkins model would be $S_n=N-C_n=N-I_n-R_n-D_n$.
Now the estimate for $R_0$ must be corrected to
$$
r_n\dfrac{N}{S_n}=r_n\dfrac{N}{N-C_n}=r_n\left(1+\dfrac{C_n}{N-C_n}\right)
$$
but this change will not be serious during the outbreak.
\biglf
If the time series $\beta_n=b_n\frac{N}{S_n}=b_n\frac{N}{N-C_n}$
for $\beta$ and $\gamma_n$ for $\gamma$
are boldly used as predictors for $\beta$ and $\gamma$ in a SIR model,
and if the model is started using $S_n=N-C_n=N-I_n-D_n-R_n$ in
the discretized form
$$
\begin{array}{rcl}
S_{n+1}-S_n&=& -\beta\dfrac{S_n}{N}I_n,      \\[0.4cm]
I_{n+1}-I_n&=& +\beta\dfrac{S_n}{N}I_n-\gamma I_n,      \\[0.4cm]
(R+D)_{n+1}-(R+D)_n&=&-\gamma I_n,
\end{array}
$$
one gets a crude prediction of the peak in case $R_0=\beta/\gamma>1$.
\biglf
Figure \RSref{FigBF} shows results for two cases. The top shows the case 
for the United States using data from day 109 (May 10$^{th}$)
and estimating $\beta$ and $\gamma$ from the data one week before.
The peak is predicted at day 472 with a total rate of 33\% Infectious.
To see how crude the technique is, the second plot shows the case
of Germany using data up to day 75, i.e. before the peak, and the peak is
predicted at day 230 with about 16\% Infected. At day 75, $R_0$
was estimated at 2.01,
but a few days later the estimate went below 1 (Figure \RSref{FigR109})
by political intervention changing $b_n$ considerably.
See Figure \RSref{FigGermany67} for a much better prediction
using data only up to day 67.
  \begin{figure}
\begin{center}
  \includegraphics[width=9.0cm,height=5.0cm]{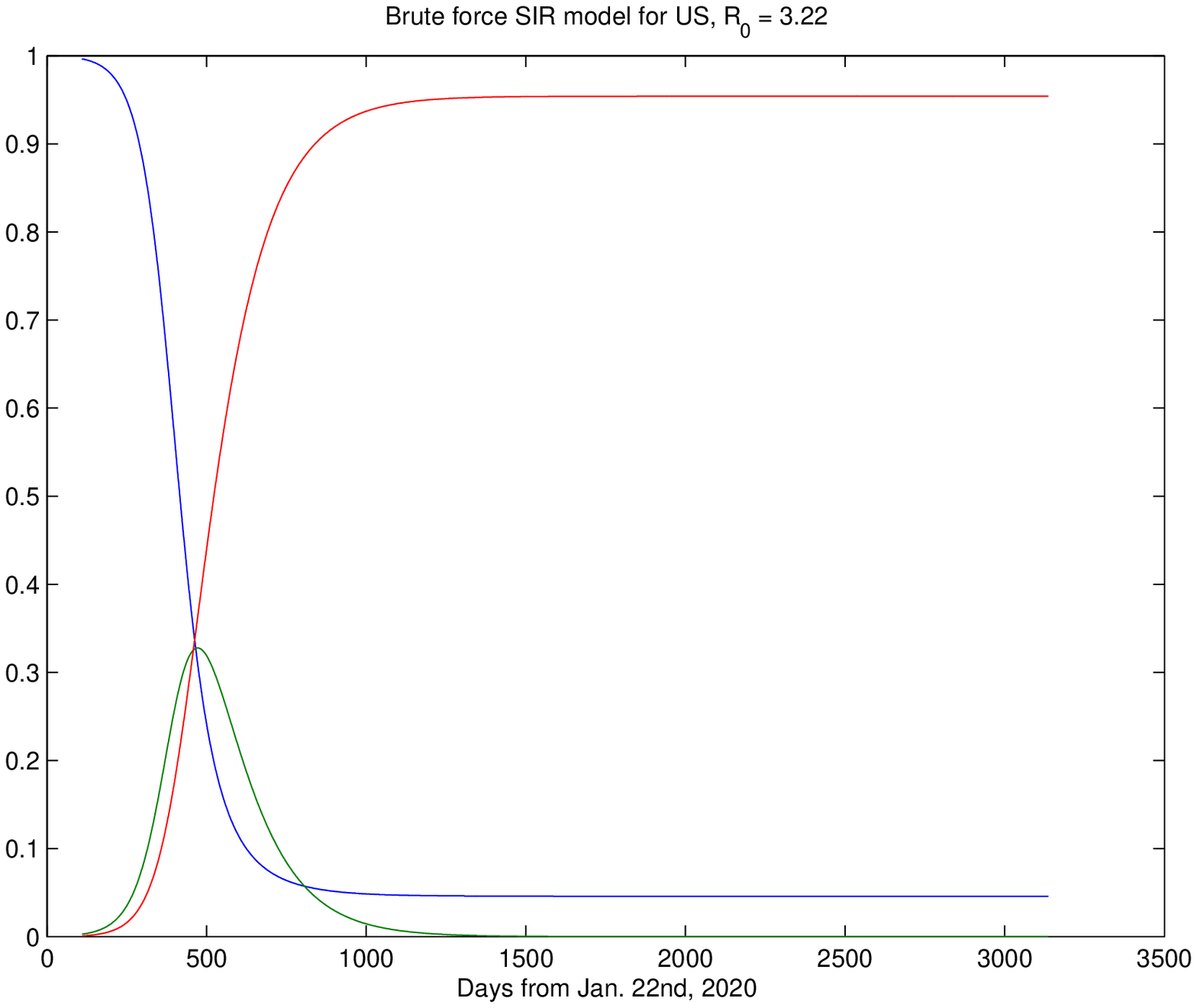}
  \includegraphics[width=9.0cm,height=5.0cm]{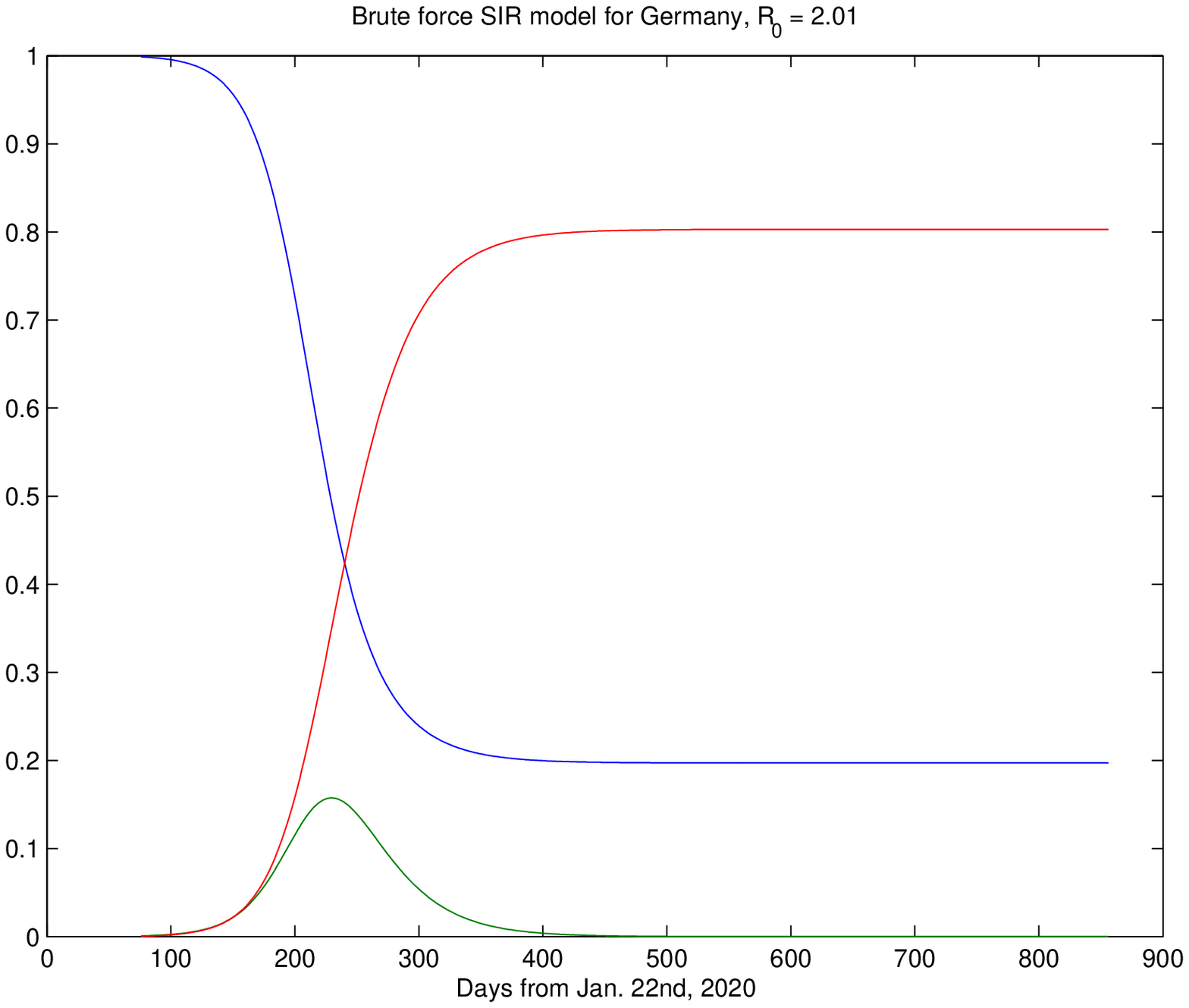}
  \caption{Brute force SIR modeling for US and Germany
    using last week's data, at days 109 and 75, respectively.
%    \red{(Program: BasicCRDI\_BruteForce.m)}
    \RSlabel{FigBF}}
\end{center} 
  \end{figure}
%****************************************************************
\section{Extended SIR Model}\RSlabel{SecEtaSIRESM}
To get closer to what actually happens,
one should combine the data-oriented Johns Hopkins Data Model
with a SIR model that accounts for what
happens outside of the Confirmed.
We introduce the time series
\begin{itemize}
\item[$S$] for the Susceptibles like in the SIR model,
\item[$M$] for the Infectious, not yet confirmed,
  ($M$ standing for {\em mysterious}), 
\item[$H$] for the unconfirmed Recovered 
  ($H$ standing for {\em healed}). 
\end{itemize}
This implies that all deaths occur within the Confirmed,
though this is a highly debatable issue. It assumes that persons with serious
symptoms get confirmed, and nobody dies of COVID-19
without prior confirmation.
%****************************************************************
\subsection{The Hidden Model}\RSlabel{SecHM}
The Removed from the viewpoint of a global SIR model
including $H$ and $M$
are $H+C$, and thus the SIR model is 
\bql{eqHidMod}
\begin{array}{rcl}
S_{n+1}-S_n&=& -\beta\dfrac{S_n}{N}M_n,      \\
M_{n+1}-M_n&=& \beta\dfrac{S_n}{N}M_n    -\gamma M_n,\\
(H+C)_{n+1}-(H+C)_n &=&  \gamma M_n.
\end{array}
\eq
To run this {\em hidden} model with constant $N=S+M+H+C$,
one needs initial values and good
estimates for $\beta$ and $\gamma$, which are not the ones
of the Johns Hopkins Data Model of section
\RSref{SecJHDM}.
%****************************************************************
\subsection{The Observable Model}\RSlabel{SecOM}
The Johns Hopkins variables $D$ and $R$ are linked to the hidden model
via $C=I-R-D$. They follow an {\em observable} model
\bql{eqObsMod}
\begin{array}{rcl}
  I_n&=&C_n-R_n-D_n,\\
  D_{n+1}-D_n&=& \gamma_{iCD} I_n,      \\
R_{n+1}-R_n&=& \gamma_{iCR} I_n
\end{array} 
\eq
with {\em instantaneous case death and recovery rates}
$\gamma_{iCD}$ and $\gamma_{iCR}$ for the
Confirmed Infectious. These rates can be estimated separately
from the available Johns Hopkins data, and we shall do this
below. We call thes rates {\em instantaneous}, because they
artificially attribute the new deaths or recoveries at day $n+1$
to the previous day, not to earlier days. In this sense, they
are rather artificial, and we shall address this question.
They are {\em case} rates, because they concern the Confirmed.  
\biglf
The observable model is coupled to the hidden model only by $C_n$.
Any data-driven $C_n$ from the observable
model can be used to enter the $H+C$ variable
of the hidden model, but in an unknown ratio.
Conversely, any version of the hidden model produces $H+C$ values
that do not determine the $C$ part. Summarizing, there is no way
to fit the hidden model to the data without additional assumptions.
\biglf
Various possibilites were tried to connect the Hidden to the
Observable. Two will be presented now.
%****************************************************************
\subsection{Infection Fatality Rates and the $k$-Day Rule}\RSlabel{SecILR}
Recall that the pa\-ra\-me\-ter $\gamma_{iCD}$ in the observable model
\eref{eqObsMod} relates case fatalities to the confirmed Infectious
of the previous day.
In contrast to this, the {\em infection fatality rate}
in the standard literature, denoted by
$\gamma_{IF}$ here,  is relating to the infection
directly, independent of the confirmation,
and gives the probability to die of
 COVID-19 after infection with the SARS-CoV-2 virus.  
 It is $\gamma_{IF}=0.56$\% by \RScite{anderheiden-buchholz:2020-1}
 and $0.66$\% by \RScite{verity-et-al:2020-1},
 but specialized for China.
Recent data from the Heinsberg study by Streeck et. al.
 \RScite{streeck:2020-1}  gives a value of
 0.37\% for the Heinsberg population in Germany.
 The idea to use the infection fatality rate for information about the hidden
 system 
 comes from \RScite{bommer-vollmer:2020-1}. The way to use
 it will depend on how to handle delays, and it turned out to
 be difficult to use these rates in a stable way.
 \biglf
Let us focus on probabilities to die either
 after an infection or after confirmation of an infection.
 The first is the infection fatality rate given in the literature,
 but what is the comparable
 case fatality rate $\gamma_{CF}$
 when using the Johns Hopkins data? It is clearly not 
 the $\gamma_{iCD}$ in \eref{eqObsMod}, giving the ratio of new
 deaths at day $n+1$ as a fraction of the confirmed Infectious at day $n$.
 It makes much more sense to use the fact that COVID-19
 diseases end after $k$ days from confirmation with either death or recovery.
 Let us call this the {\em $k$-day rule}.
 Suggested values for $k$ range between 14 and 21.
 %****************************************************************
\subsubsection{Estimation of Case Fatality Rates}\RSlabel{SecEoCFR}
 Following 
 \RScite{schaback:2020-3}, one can estimate the probability $p_i$
 to die on day $i$ after confirmation, and this works in a stable way
 per country,
 based only on $C$ and $D$, not on the unstable $R$
 data. In \RScite{schaback:2020-3}
 this approach was used to produce $R$ values that comply with the $k$-day
 rule, but here we use it for estimating
 the case fatality. 
 The technique of \RScite{schaback:2020-3} performs a fit
 \bql{eqDDpCC}
 \begin{array}{rcl}
   D_n-D_{n-1}&\approx& \displaystyle{\sum_{i=1}^kq_i(C_{n-i}-C_{n-i-1})   },\\
   q_i&=&p_i \displaystyle{\prod_{j=0}^{i-1}(1-p_j),\;1\leq i\leq k+1}
 \end{array} 
\eq
i.e. it assigns all new deaths at day $n$ to previous new infections
on previous days in a hopefully consistent way, minimizing the error
in the above formula under variation of the probabilities $p_i$.
If the $p_i$ are known for days 1 to $k$, the
{\em case fatality rate}
is
$$
\gamma_{CF}=\displaystyle{  \sum_{i=1}^k q_i=1-q_{k+1}.}
$$
This argument can also be seen the other way round:
the new Confirmed $C_{n}-C_{n-1}$ at day 1 enter into $D_{n+1}-D_n$
with probability $q_1=p_1$, into $D_{n+2}-D_{n+1}$ with probability
$q_2=p_2(1-p_1)$ and so on. The rest enters into
the new Recovered at day $n+k$ with probability $q_{k+1}$ if we set
$p_{k+1}=0$. Thus the case fatality rate can be expressed as $1-q_{k+1}$
like above.
\biglf
At this point, there is a hidden assumption. Persons that are new
to the Confirmed at day $n$ are not dead and not recovered.
The change $C_{n+1}-C_n$ to the Confirmed is therefore understood as
the number of new registered infections. Otherwise, one might
replace $C_{n-i}-C_{n-i-1}$ by $I_{n-i}-I_{n-i-1}$ in \eref{eqDDpCC},
but this would connect a cumulative function to a non-cumulative function.
Furthermore, this would use the unsafe data
of the Recovered. 
\biglf
In fact, equation \eref{eqDDpCC} is unexpectedly reliable,
provided one looks at the sum of the probabilities $p_i$.
This follows from  series of experiments that we do not document fully here,
In \RScite{schaback:2020-3}, data for
 $2k$ days backwards were used for the estimation, and
 results did not change much when more or less data were used
 or when $k$ was modified.
 Here, the range $7\leq k\leq 21$ was tested, and
 backlogs of up to 50 days from day 109 (as of this writing).
 See Figure \RSref{FigcaseDRItaly109} below for an example. 
 Larger $k$ lead to somewhat larger fatality rates, because the method has more
 leeway to assign deaths to confirmations, but the increase is rather small when
 $k$ is increased beyond 14. It is remarkable that $k=7$ suffices for a rather
 good fit for US, UK, Sweden, and Italy. In contrast to other countries, this
 means that 7 days after confirmation are enough to assign deaths
 consistently to previous confirmations, indicating that a large portion
 of confirmations concern rather serious cases.
 \biglf
 In general, the fit gets better when
 the backlog is not too large, avoiding use of outdated data, and the resulting
 rate does not change much when backlogs are shortened.
 Roughly, the rule of a backlog of $2k$ days works well, together with
 $k=21$ to allow enough leeway to ensure a small fitting error. Note that all
 variations of $k$ and the backlog data have a rather small effect on the sum of
 the $p_i$, while the $p_i$ will vary strongly.
 \biglf
 See the first column of Table \RSref{TabCLR}
 for estimates of case fatality rates for different countries,
 calculated on day 109 (May 10$^{th}$). These depend strongly on the
 strategy for confirmation. In particular, they are high when
 only serious cases are confirmed, e.g. cases that need hospital care.
 If many more people are tested, confirmations will contain
 plenty of much less serious cases, and then the case fatality rates
 are low. The values were entered manually after inspection of
 a plot of the rates as functions of $k$ and the backlog. See Figure
 \RSref{FigcaseDRItaly109} for how the value 0.143
 for Italy was determined.

 \begin{figure}
 \begin{center}
  \includegraphics[width=12.0cm,height=7.0cm]{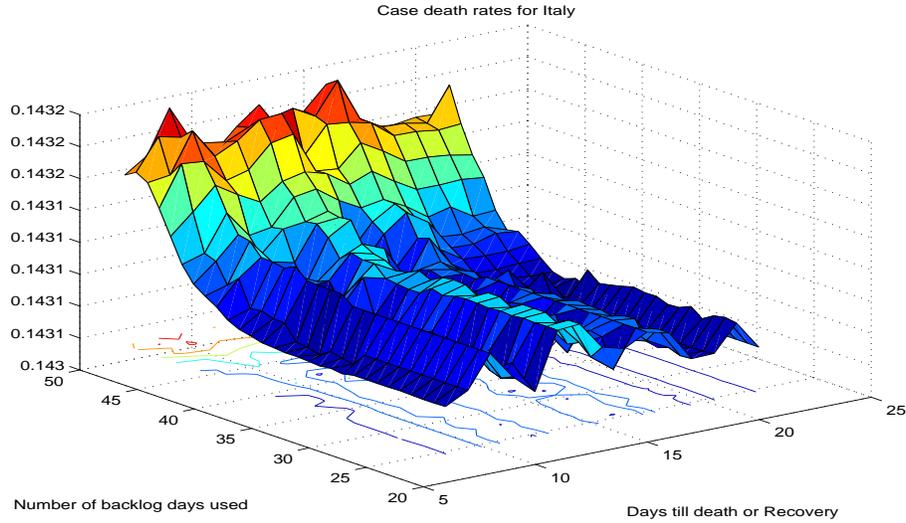}
  \caption{Case fatality rate
    for Italy based on data at day 109,
    as functions of $k$ and the data backlog
    %\red{(Program: BasicCRDIDRRates.m)}
    \RSlabel{FigcaseDRItaly109}}
\end{center} 
\end{figure} 

 % \red{(Program: BasicCRDIDRRates.m)}
 \begin{table}
   \begin{center}
     \begin{tabular}{||r||r|r|}
       \hline\hline
       Country & Death & Detection\\
               & rate  & rate\\
       \hline\hline
Germany   & 0.046  & 0.130\\   
South Korea   & 0.020 & 0.300\\   
Italy     & 0.143  & 0.042\\   
Spain     & 0.117  & 0.051\\   
Sweden    & 0.157  & 0.038\\   
Austria   & 0.040  & 0.150 \\   
France    & 0.151 & 0.040 \\   
UK        & 0.157  & 0.038\\   
US        & 0.067  & 0.089 \\
\hline\hline
\end{tabular} 
     \caption{Case fatality and detection rates, estimated on day 109
       % \red{(Program: BasicCRDIDRRates.m)}
       \RSlabel{TabCLR}}
   \end{center}
 \end{table}
\biglf
The instantaneous case death
rate $\gamma_{iCD}$ of \eref{eqObsMod} for the Johns Hopkins
 data comes out around 0.004 for Germany
 by direct inspection of the data via
 \bql{eqDDIg}
\gamma_{iCD}\approx \dfrac{D_{n+1}-D_n}{I_n},
 \eq
while the Case Fatality Rate $\gamma_{CF}$
in Table \RSref{TabCLR} is about 0.045. The deaths have to be attributed to
different days using the $k$-day rule, they cannot easily be
assigned to the previous day without making the rate smaller.
%****************************************************************
\subsubsection{Using Fatality Rates for the Hidden Model}\RSlabel{SecUFR}
 If the case fatality rates  $\gamma_{CF}$ of 
 Table \RSref{TabCLR} are used with the infection fatality rate of
 $\gamma_{IF}=0.006$,
 one should obtain an estimate of the total Infectious.
 If the formula \eref{eqDDpCC} is written as
 $$
 \displaystyle{\sum_{i=1}^kq_i(C_{n-i}-C_{n-i-1})   }\approx
 D_n-D_{n-1}\approx \displaystyle{\sum_{i=1}^k\tilde q_i(S_{n-i-1}-S_{n-i})   }
 $$
in terms of the previous new infections $S_{n-i-1}-S_{n-i}$ with
infection fatality probabilities $\tilde q_i$, one should maintain
$$
\gamma_{CF}=\displaystyle{\sum_{i=1}^k q_i} \hbox{ and } \displaystyle{  
  \gamma_{IF}=\sum_{i=1}^k \tilde q_i   } 
$$
and this works by setting
\bql{eqCCggSS}
C_{n}-C_{n-1}=\dfrac{\gamma_{IF}}{\gamma_{CF}}   (S_{n-1}-S_{n})
\eq
in general, without using the unstable $p_i$.
%****************************************************************
\subsubsection{The Detection Rate}\RSlabel{SecDR}
The quotient $\frac{\gamma_{IF}}{\gamma_{CF}}$ can be called
the {\em detection rate}, stating the fraction of Infectious that is
entering confirmation. See the second column of Table \RSref{TabCLR}.
Note that all of this is dependent on good estimates of the
infection fatality rate, and the new value by \RScite{streeck:2020-1}
will roughly double the detection rate for Germany. 
\biglf
All of this is comparable to the findings of
\RScite{bommer-vollmer:2020-1} and uses the basic
idea from there, but is executed with a somewhat different technique.
\biglf
A simple way to understand the quotient $\frac{\gamma_{IF}}{\gamma_{CF}}$
as a detection rate is to ask for the probability $p(C)$ for Confirmation.
If the probability to die after Confirmation is $\gamma_{CF}$, and if there are
no deaths outside confirmation, then
$$
p(D)=p(C)\cdot p(D|C),
$$
and
$$
p(C)=\dfrac{p(D)}{p(D|C)}=\dfrac{\gamma_{IF}}{\gamma_{CF}}.
$$
It is tempting to replace $S$ by $M$ in \eref{eqCCggSS}, but
this would make $M$ cumulative. 
%****************************************************************
\subsubsection{Local Estimation of Fatality Rates}\RSlabel{SecLEoFR}
Under political changes of the pa\-ra\-me\-ters $\beta$ and $\gamma$,
the estimation of the Case Fatality Rate should be made locally, not globally.
Using the experience of \RScite{schaback:2020-3} and section \RSref{SecEoCFR},
we shall do this using a fixed $k=21$ for the $k$-day rule
and data for a fixed backlog of $2k$ days. 
%****************************************************************
\subsection{Recovery Rates}\RSlabel{SecRR}
We need another pa\-ra\-me\-ter to make the model work.
There are many choices, and after some failures we selected the
constant $\gamma_{iIR}$ in a model equation
$$
H_{n+1}-H_n=\gamma_{iIR}M_n.
$$
Following what was mentioned about {\em instantaneous}
rates in section \RSref{SecOM}, this
is an {\em instantaneous Infection Recovery Rate},
relating the new unregistered Recovered to the
unregistered Infections the day before.
%****************************************************************
\subsubsection{Estimation of Recovery Rates}
A good value of $\gamma_{iIR}$ can come out of an experiment
that produces a time series for $M$ and $H$, i.e. for unregistered Infectious
and unregistered Recovered. Then the
instantaneous Infection Recovery rate $\gamma_{iIR}$
can be obtained directly by
$$
\dfrac{H_{n+1}-H_n}{M_n}\approx\gamma_{iIR}.
$$
The Infection Recovery rate $\gamma_{IR}=1-\gamma_{IF}$
does not help much, because we need an instantaneous rate
that has no interpretation as a probability.
\biglf
Without additional input, we can look at the
instantaneous Case Recovery rate
\bql{eqRRIg}
\dfrac{R_{n+1}-R_n}{I_n}\approx\gamma_{iCR}
\eq
that is available from the Johns Hopkins data, and comes out experimentally
to be rather stable. The rate $\gamma_{iIR}$ must be larger, because
we now are not in the subpopulation of the Confirmed, and nobody can die
without going first into the population of the Confirmed. As long as no better
data are available, we use the formula
\bql{eqggg}
\gamma_{iIR}=\dfrac{1-\gamma_{IF}}{1-\gamma_{CF}}\gamma_{iCR}=
\dfrac{\gamma_{IR}}{\gamma_{CR}}\gamma_{iCR}=\dfrac{\gamma_{iCR}}{\gamma_{CR}}\gamma_{IR}
\eq
that accounts for two things:
\begin{enumerate}
 \item the value $\gamma_{iCR}$ is increased by the ratio of Recovered
   probabilities
   for the Infected and the Confirmed,
 \item the value $\gamma_{IR}$ is multiplied by a factor for 
   transition to immediate rates, and this factor is the
   transition factor for the Confirmed Recovered. 
\end{enumerate}
The above strategy is debatable and may be the weakest point of this
approach. However, others turned out to be worse, mainly due
to instability of results. Furthermore, the final prediction does not need
it. It enters only in the intermediate step when $S,\,M$, and $H$ are calculated
in the time
range of the available Johns Hopkins data, see section
\RSref{SecModComp}. And, finally, there is hope that
there will be medical experiments that yield reliable values directly.
%****************************************************************
\subsubsection{Practical Approximation of Recovery Rates}\RSlabel{SecPAoRR}
In \eref{eqggg} the rate $\gamma_{IR}$ is fixed, and the rate $\gamma_{CR}$
is determined locally via section \RSref{SecLEoFR}. The rate
$\gamma_{iCR}$ follows from the time series
$$
\dfrac{R_{n+1}-R_n}{I_n}\approx \gamma_{iCR}
$$
as in \eref{eqObsMod}. This works for countries that provide useful
data for the Recovered. In that case, and in others to follow below,
we can take the time series itself as long as we have data.
For prediction,
we estimate the constant from the time series using a fixed backlog of $m$
days from the current day. Since many data have a weekly oscillation,
due to data being not properly delivered during weekends,
the backlog should not be less than 7.
\biglf
But for certain countries, like the United Kingdom, the
data for the Recovered are useless. In such cases, we employ the
technique of \RScite{schaback:2020-3} to estimate the Recovered
using the $k$-day rule and a backlog of $2k$ days, like in
section \RSref{SecLEoFR} for the case fatality rates. 
%****************************************************************
\subsection{Model Completion}\RSlabel{SecModComp}
We now have everything to run the hidden model, but we do it first
for days that have available Johns Hopkins data. This leads to
estimations of $S,\,M$, and $H$
from the observed data of the Johns Hopkins source.
With the pa\-ra\-me\-ters from above, we use the new relations 
\bql{eqCCHHggSS} % eqCCggSS
\begin{array}{rcl}
  C_{n+1}-C_{n}&=&\dfrac{\gamma_{IF}}{\gamma_{CF}}   (S_{n}-S_{n+1}),\\
  H_{n+1}-H_{n}&=&\gamma_{iIR}M_{n}.\\
\end{array}
\eq
The first equation is a priori and determines $S$. One could  run it over
the whole range of available data, if the pa\-ra\-me\-ter $\gamma_{CF}$
were fixed, like $\gamma_{CI}$.
But since we estimate it via section \RSref{SecILR}
that resulted in Table \RSref{TabCLR}, we should use
section \RSref{SecLEoFR} to calculate $\gamma_{CF}$ locally.
\biglf
The second is a posteriori
and lets $H$ follow from $M$, but we postpone it.
The third model equation in \eref{eqHidMod} will
be handled defining $\gamma_n$ by brute force via
\bql{eqgMCCHH}
\begin{array}{rcl}
  \gamma_n M_n
  &=&
  C_{n+1}-C_{n}+H_{n+1}-H_{n}.
\end{array}
\eq
We then set up the second model equation for $M$ as
\bql{eqMMSSN}
\begin{array}{rcl}
  M_{n+1}-M_n
  &=&
  S_n-S_{n+1}-\gamma_n M_n\\
  &=&
  \dfrac{\gamma_{CF}}{\gamma_{IF}}( C_{n+1}-C_{n})-\gamma_n M_n\\
  &=&
  \dfrac{\gamma_{CF}}{\gamma_{IF}}( C_{n+1}-C_{n})-(C_{n+1}-C_{n}+H_{n+1}-H_{n})\\
  &=&
  \left(\dfrac{\gamma_{CF}}{\gamma_{IF}}-1\right)( C_{n+1}-C_{n})-\gamma_{iIR}M_{n}\\
\end{array}
\eq
that can be solved if an initial value is prescribed. The first model equation
in \eref{eqHidMod} is used to define $\beta_n$ by
\bql{eqSSbSNM}
\begin{array}{rcl}
S_{n}-S_{n+1}&=&\beta_n\dfrac{S_n}{N}M_n.
\end{array}
\eq
The model is run by executing \eref{eqMMSSN} for a starting value $\tilde M_0$,
yielding the $M_n$. Then \eref{eqCCHHggSS}
is run to produce the $S_n$ and $H_n$,
with starting values. 
If $\beta_n$ and $\gamma_n$ are calculated by \eref{eqSSbSNM} and
\eref{eqgMCCHH},
respectively, the balance equation $N=S+M+H+C$
follows from \eref{eqMMSSN}.
%****************************************************************
\subsubsection{Influence of Starting Values}\RSlabel{SecIoSS}
Since the populations are large, the starting values
for $S$ are not important. The starting value for $H$ is irrelevant
for $H$ itself, because only
differences enter, but it determines the  starting value for $M$ due to
the balance equation. Anyway, it turns out experimentally that the starting
values do not matter, if the model is started early. The hidden model
\eref{eqHidMod} depends much more strongly on $C$ than on the
starting values.  See Figure \RSref{FigGermany67} for an example.
%\biglf
%\red{Program BasicCRDIModelCH.m}
%****************************************************************
\subsubsection{Parameter Estimation}\RSlabel{SecModCompPE}
Along with the calculation of $S,\,M$, and $H$, we run the calculation
of the time series $\beta_n$ and $\gamma_n$ using
\eref{eqSSbSNM} and \eref{eqgMCCHH}. These yield estimates for the
pa\-ra\-me\-ters of the full SIR
model that replace the earlier time series from the
Johns Hopkins Data Model in section \RSref{SecJHDM}.
%****************************************************************
\subsubsection{Examples}\RSlabel{SecModCompEx}
The figures to follow in section \RSref{SecExaFull}
show the original
Johns Hopkins data together with the hidden variables $S,\,M$, and $H$
that are calculated by the above technique. Note that the only
ingredients beside the Johns Hopkins data are the number $k$
for the $k$-day rule, the Infection Fatality Rate $\gamma_{IF}$
from the literature, and the backlog $m$ for estimation of constants
from time series. 
%****************************************************************
\section{Predictions using the Full Model}\RSlabel{SecPutFM}
To let the combined model predict the future, or to check
what it would have predicted if used at an earlier day,
we take the completed model of the previous sections up to a day
$n$ and use the values $S_n,\,M_n,\,H_n,\,C_n,\,I_n,\,R_n$ and $D_n$
for starting the prediction. With the variable $HC:=H+C$,
we   use  the recursion
\bql{eqFullMod}
\begin{array}{rcl}
  S_{i+1}&=& S_i -\beta\dfrac{S_i}{N}M_i,\\[0.4cm]
  M_{i+1}&=& M_i +\beta\dfrac{S_i}{N}M_i-\gamma M_i,\\[0.4cm]
  HC_{i+1}&=& HC_i+\gamma M_i,\\
  C_{i+1}&=& C_i +\gamma_{IF}(S_i-S_{i+1})/\gamma_{CF},\\
  R_{i+1}&=& R_i +\gamma_{iCR}I_i,\\
  D_{i+1}&=& D_i +\gamma_{iCD}I_i,\\
  I_{i+1}&=& C_{i+1}-R_{i+1}-D_{i+1},\\
  H_{i+1}&=& HC_{i+1}-C_{i+1}.
\end{array}
\eq
This needs fixed values of $\beta$ and $\gamma$ that we
estimate from the time series for $\beta_n$ and $\gamma_n$ by
using a backlog of $m$ days, following Section \RSref{SecModCompPE}.
The instantaneous rates $\gamma_{iCR}$ and $\gamma_{iCD}$ can be
calculated via their time series, as in \eref{eqRRIg} and \eref{eqDDIg},
using the same backlog. 
We do this at the starting point of the prediction, and then the model runs
in what can be called a {\em no political change} mode. Examples
will follow in section \RSref{SecExaFull}.
%****************************************************************
\subsection{Properties of the Full Model}\RSlabel{SecProtFM}
The first part of the full model \eref{eqFullMod} runs as a standard SIR model
for the variables $S,\,M$ and $H+C$, and inherits the properties of
these as described in section \RSref{SecClasSIRMod}. It does not
use the $\gamma_{iIR}$ pa\-ra\-me\-ter of the second equation
in \eref{eqCCHHggSS}, and it uses the first the other way round, now
determining $C$ from $S$, not $S$ from $C$. 
\biglf
The balances $N=S+M+H+C$ and $C=I+D+R$ are maintained automatically,
and the time series for $S,\,C,\,R$, and $D$ stay monotonic
as long as $M$ and $I$ are nonnegative. To check the monotonicity of $H$,
consider
$$
\begin{array}{rcl}
  H_{i+1}-H_i &=& HC_{i+1}-HC_i-C_{i+1}+C_i\\
  &=& \gamma M_i-\dfrac{\gamma_{IF}}{\gamma_{CF}}(S_i-S_{i+1})\\
  &=& \left(\gamma -\beta\dfrac{\gamma_{IF}}{\gamma_{CF}}\dfrac{S_i}{N}
  \right)M_i.
\end{array}
$$
The bracket is positive if
$$
R_0=\dfrac{\beta}{\gamma}<\dfrac{\gamma_{CF}}{\gamma_{IF}}\dfrac{N}{S_i}
\geq \dfrac{\gamma_{CF}}{\gamma_{IF}},
$$
which is enough for practical purposes.
%****************************************************************
\subsection{Examples of Predictions}\RSlabel{SecExaFull}
%****************************************************************
Figure \RSref{FigSMH109} shows predictions on day 109, May 10th,
for Germany, Sweden, US and UK, from the top. The plots for countries
behind their peak are rather similar to the one for Germany.
The other three countries are selected because they still have to face
their peak, if no action is taken to change the pa\-ra\-me\-ters.
The estimated $R_0$ values are 0.59, 1.37, 3.34, and 1.20, respectively.
Note that these are not directly comparable to
Figure \RSref{FigR109}, because they are the fitted constant to
the backlog of a week, and using \eref{eqgMCCHH} and \eref{eqSSbSNM}
instead of \eref{eqrbg}.
The black and magenta curves are the estimated $M$ and $H$ values, while the
$S$ values are hardly visible on the top. The hidden $M$ and $H$ in
black and magenta follow roughly the observable $I$ and $C$ in
blue and cyan, but with a factor due to the detection rate
that is different between countries, see Table \RSref{TabCLR}. 
\begin{figure}
 \begin{center}
  \includegraphics[width=12.0cm,height=5.0cm]{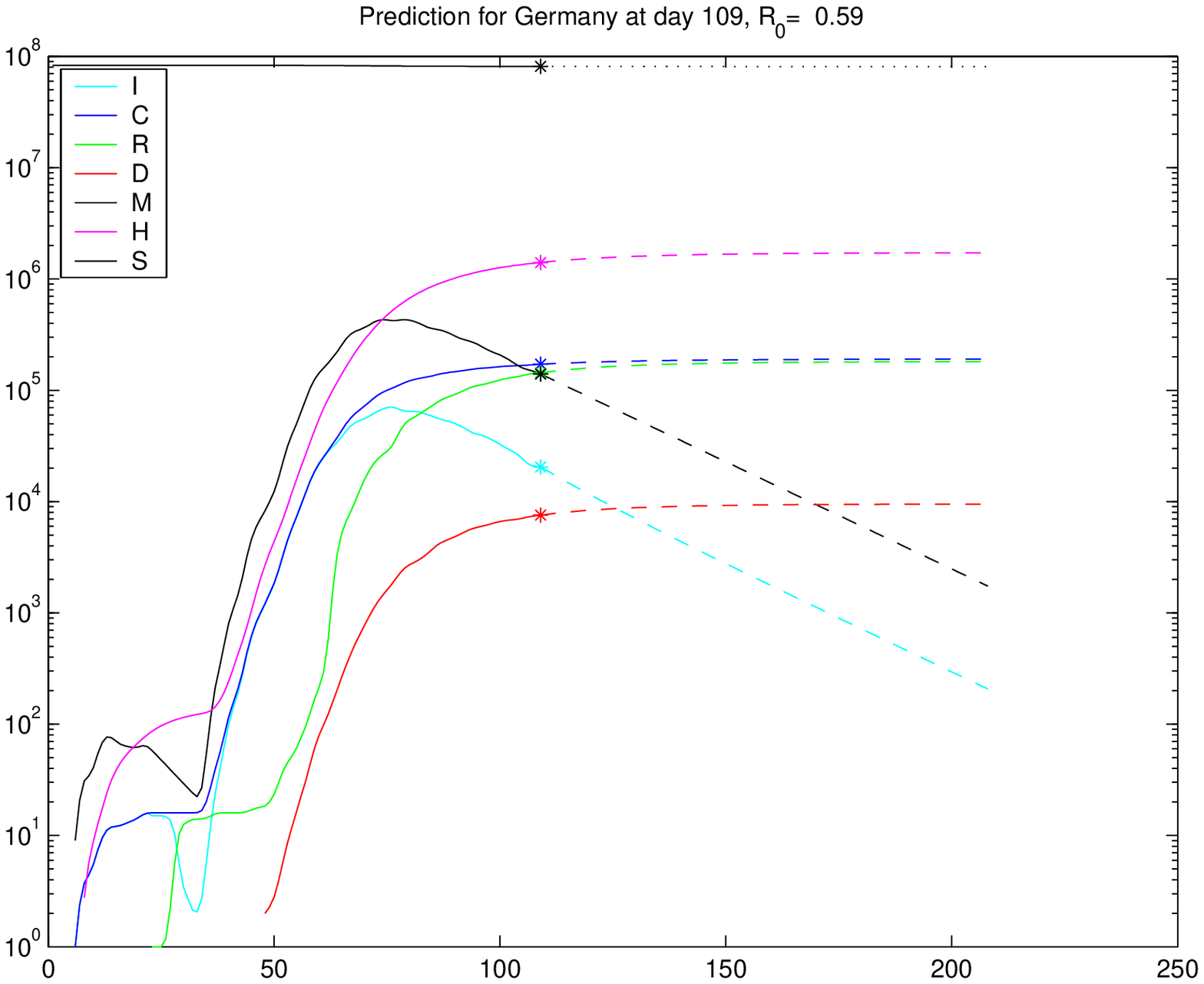}\\
  \includegraphics[width=12.0cm,height=5.0cm]{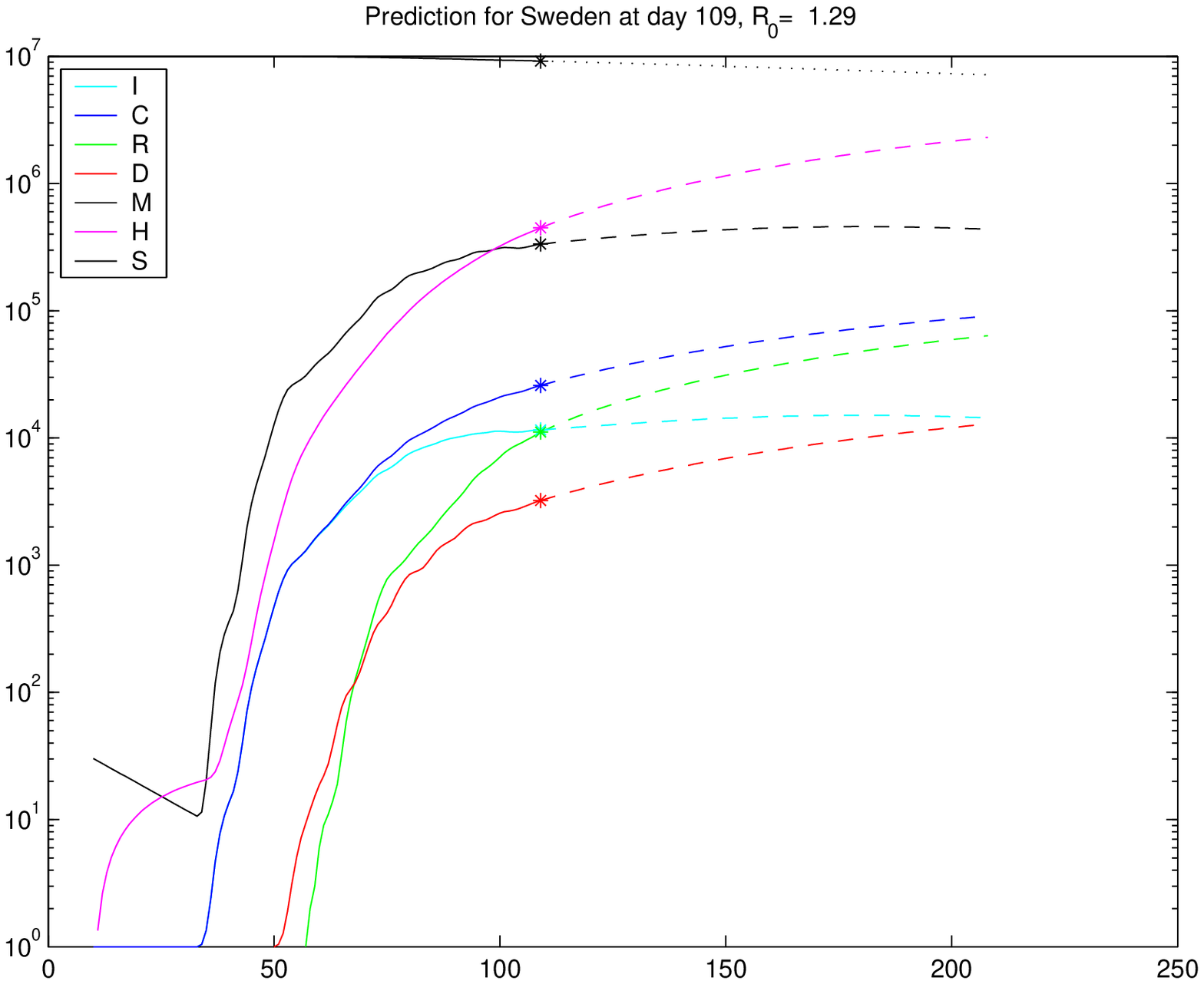}\\
  \includegraphics[width=12.0cm,height=5.0cm]{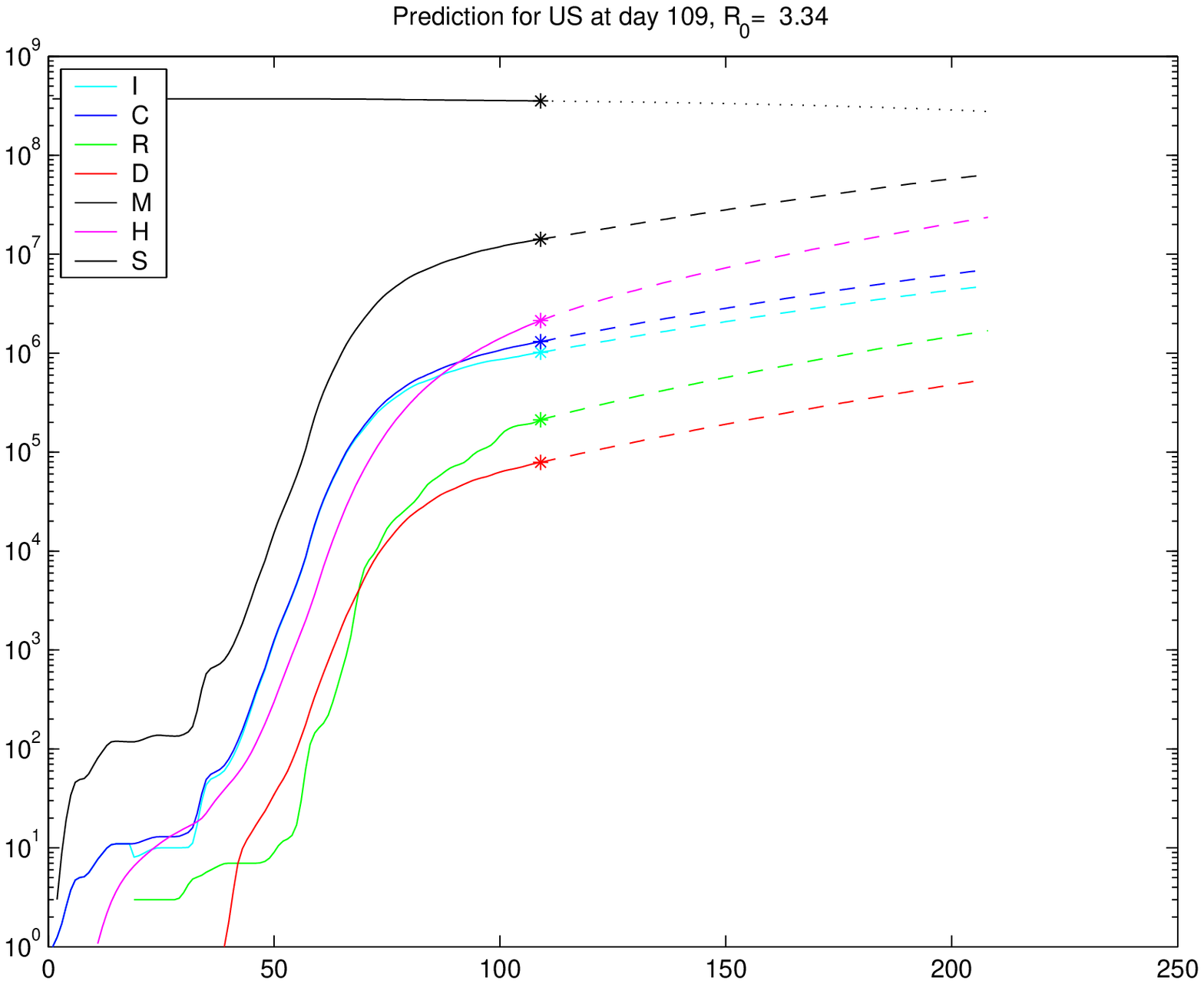}\\
  \includegraphics[width=12.0cm,height=5.0cm]{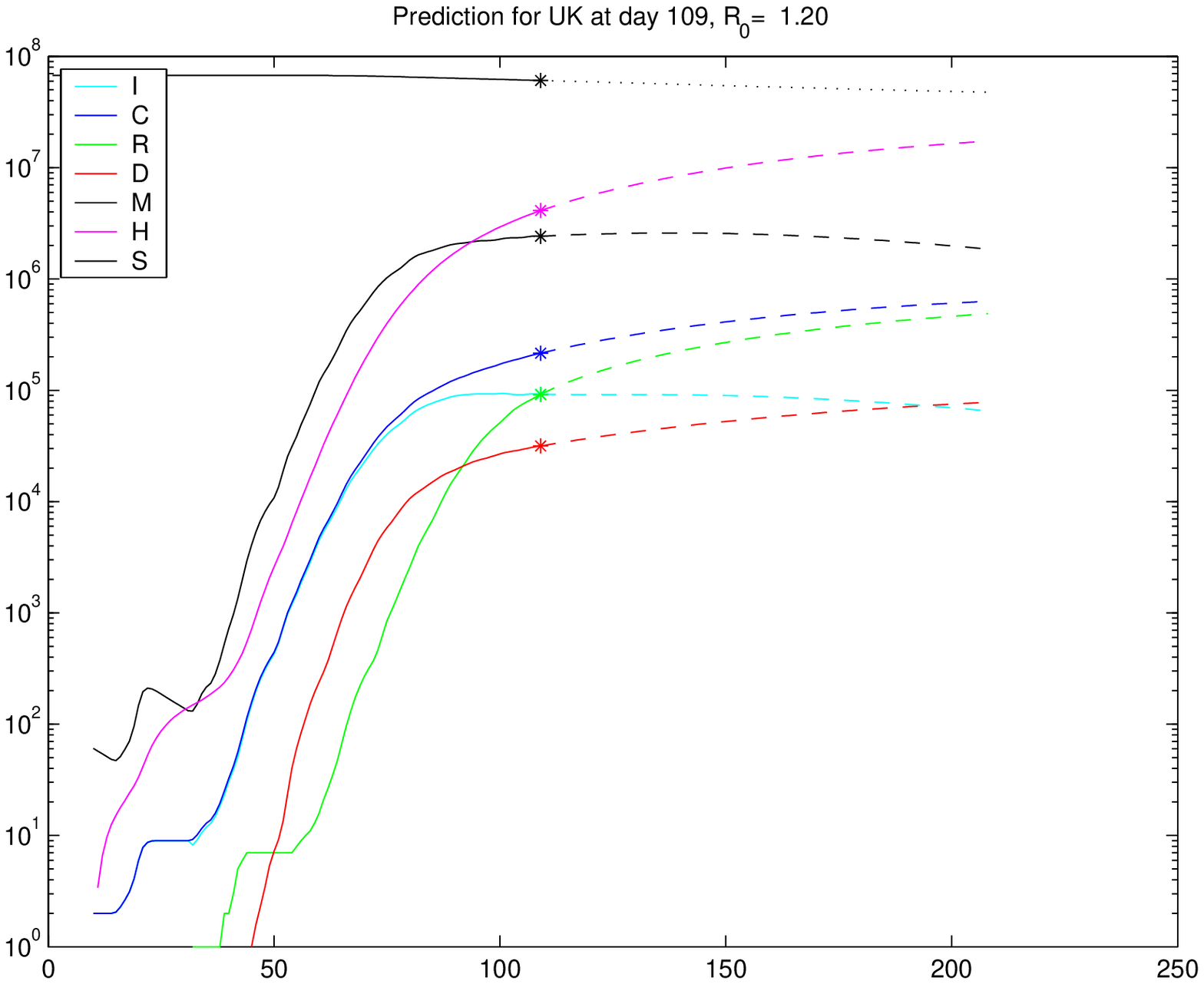}
  \caption{Predictions for countries Germany, Sweden, US, and UK on day 109
    \RSlabel{FigSMH109}} 
\end{center} 
\end{figure}
\biglf
To evaluate the predictions, one should go back and start the predictions
for earlier days, to compare with what happened later. Figure
\RSref{FigSMH109b3} shows overplots of predictions
for days 81, 95, and 109, each a fortnight apart.
The starting points of the predictions are marked by asterisks again.
Now each prediction has slightly different estimates for $S,\,M$, and $H$
due to different available data. Recall that the
determination of these variables
is done while there are Johns Hopkins data available, following section
\RSref{SecModComp}, and will be dependent on the data-driven estimations
described there. In particular, the case fatality rates
and detection rates 
of Table \RSref{TabCLR} change with the starting point of the prediction,
and they determine $S,\;M$, and $H$ backwards, see the figure for Sweden.
\begin{figure}
 \begin{center}
  \includegraphics[width=12.0cm,height=5.0cm]{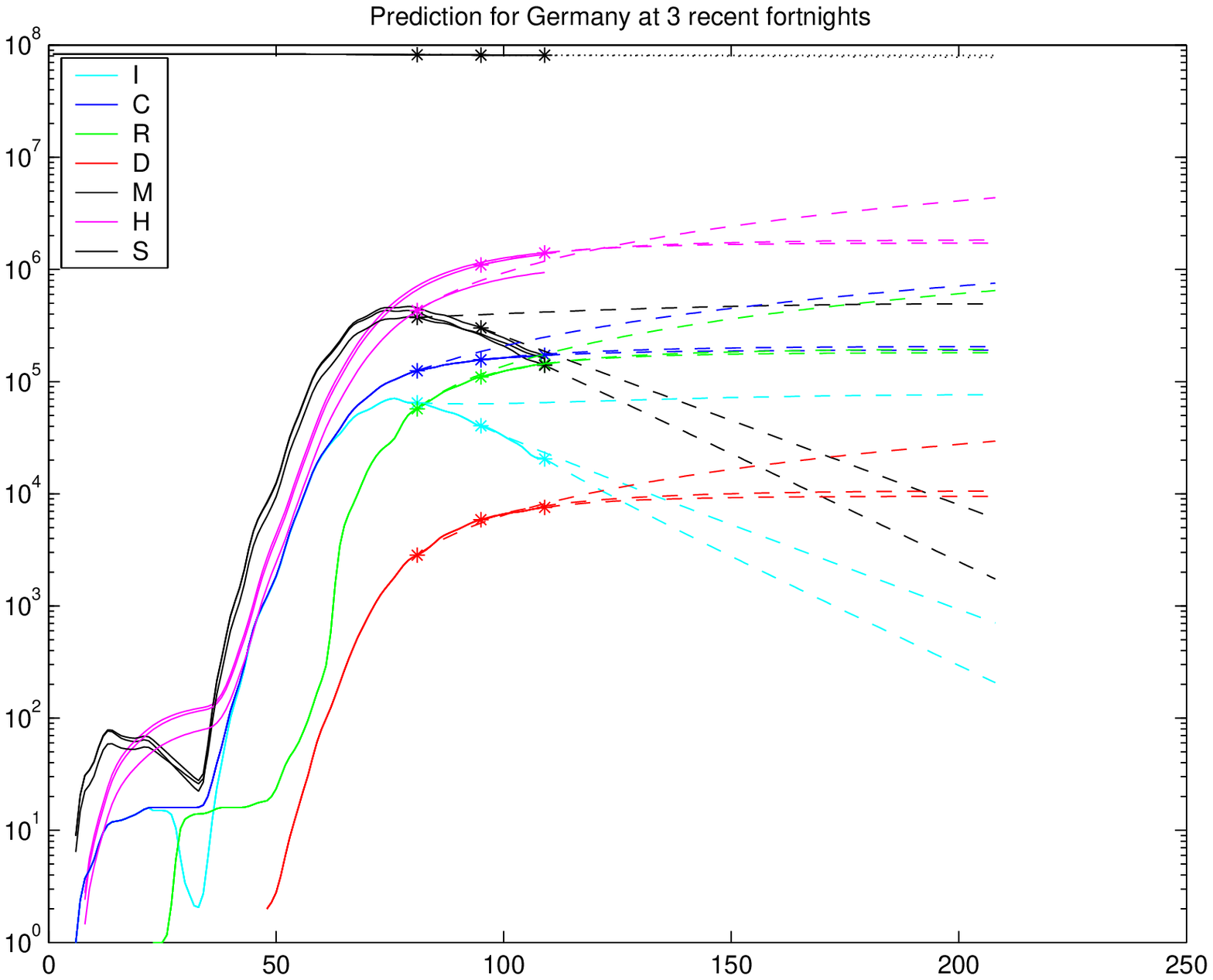}\\
  \includegraphics[width=12.0cm,height=5.0cm]{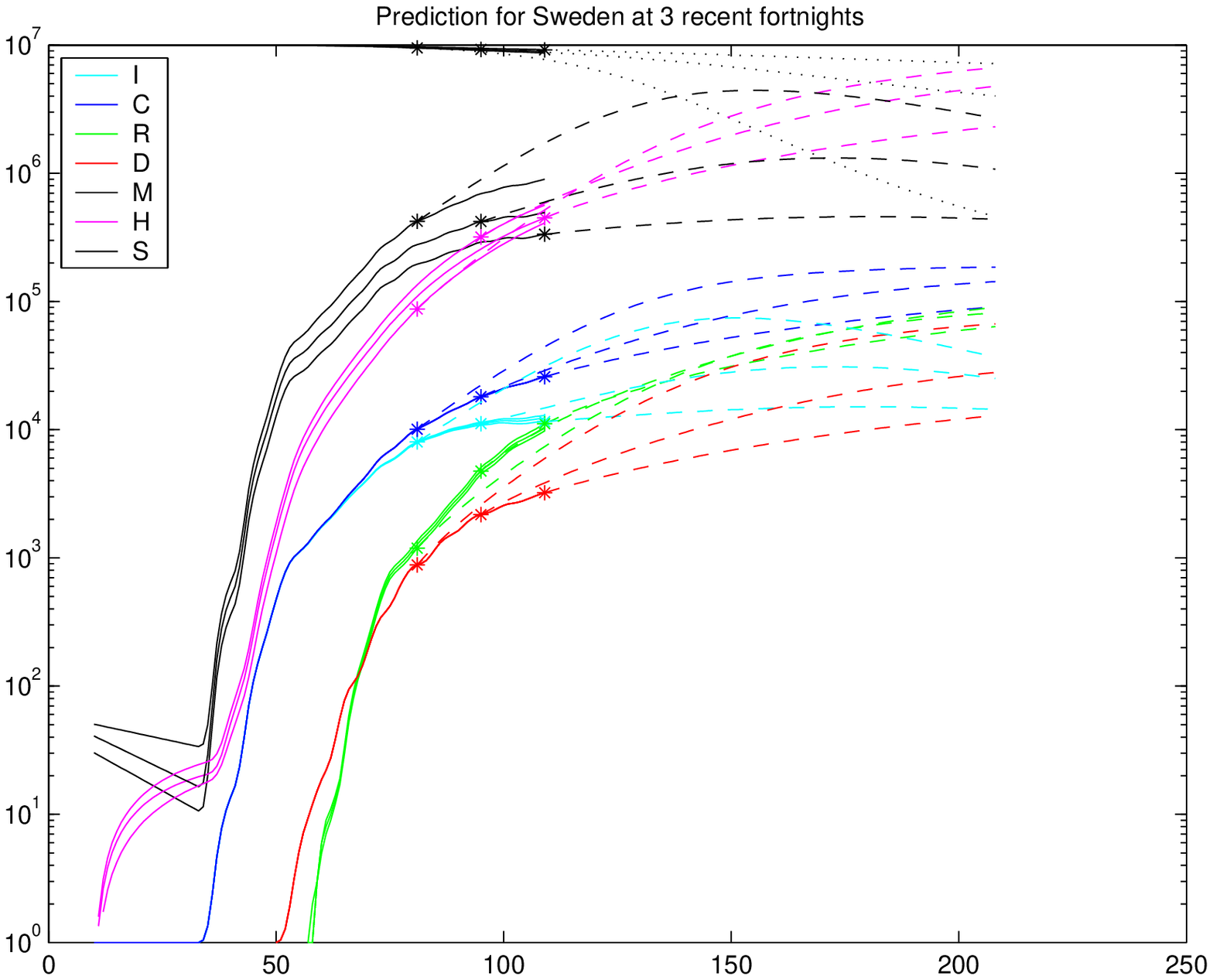}\\
  \includegraphics[width=12.0cm,height=5.0cm]{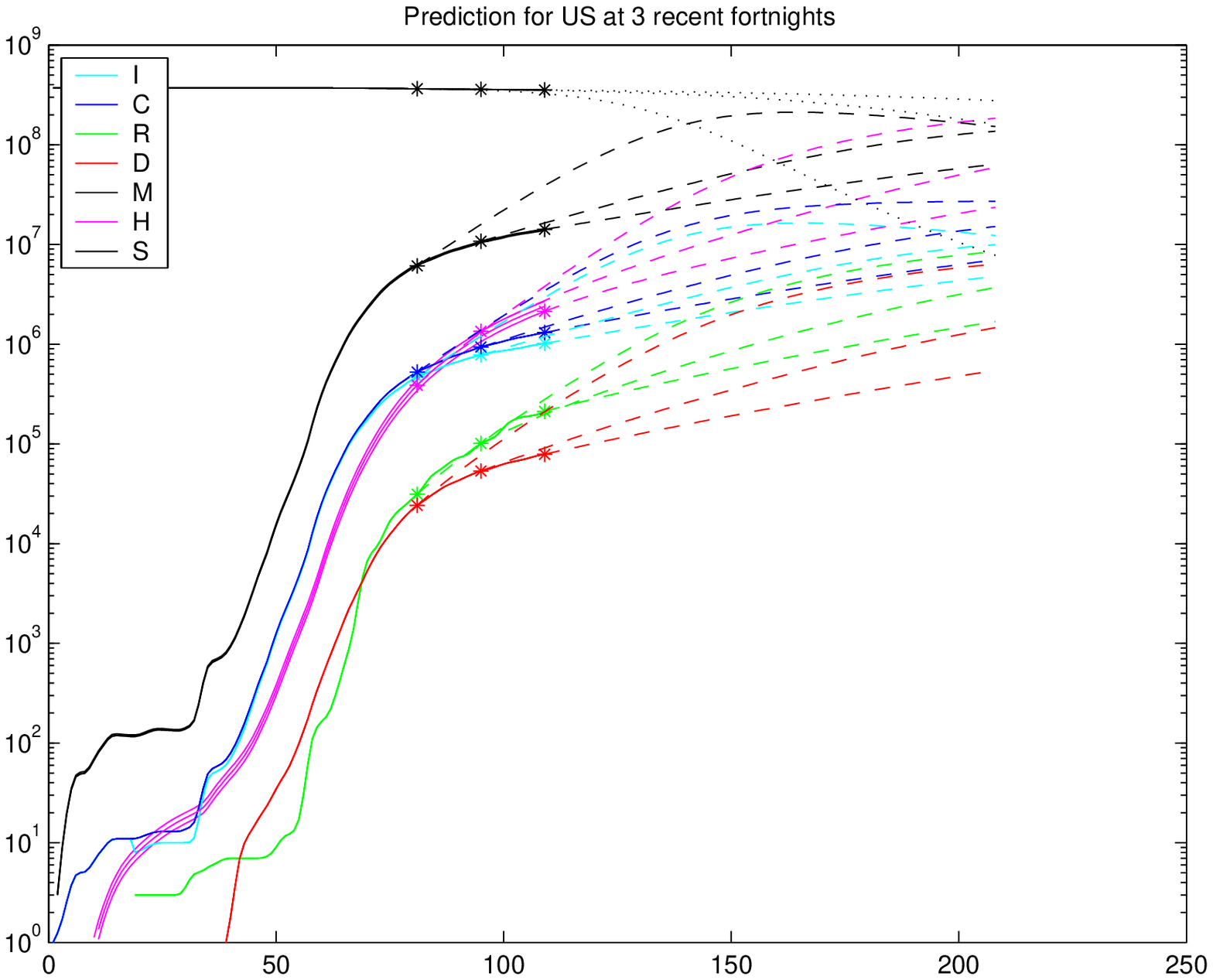}\\
  \includegraphics[width=12.0cm,height=5.0cm]{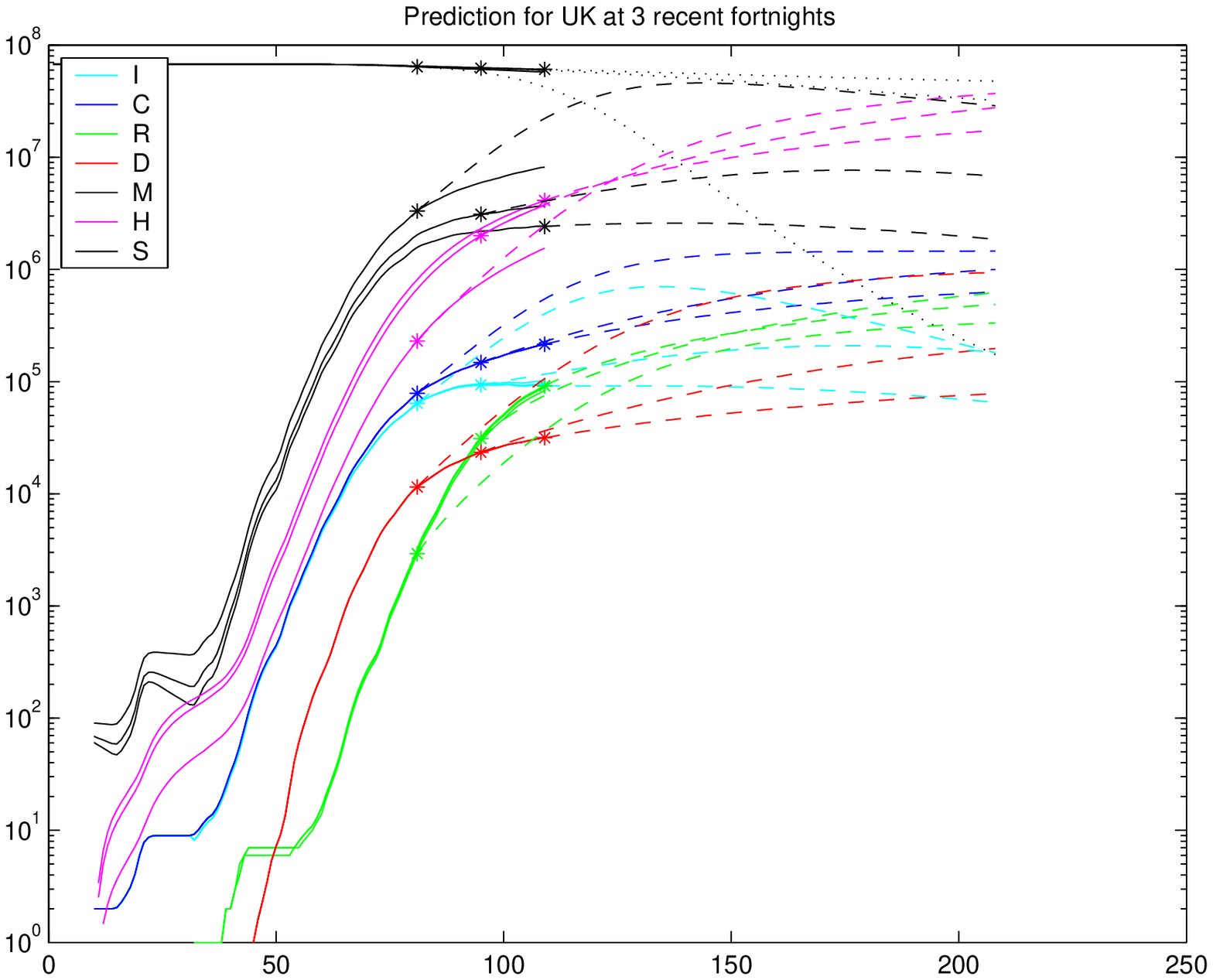}
  \caption{Predictions for countries Germany, Sweden, US, and UK on days
    109, 95, and 81
    \RSlabel{FigSMH109b3}} 
\end{center} 
\end{figure}
\begin{figure}
 \begin{center}
  \includegraphics[width=12.0cm,height=5.0cm]{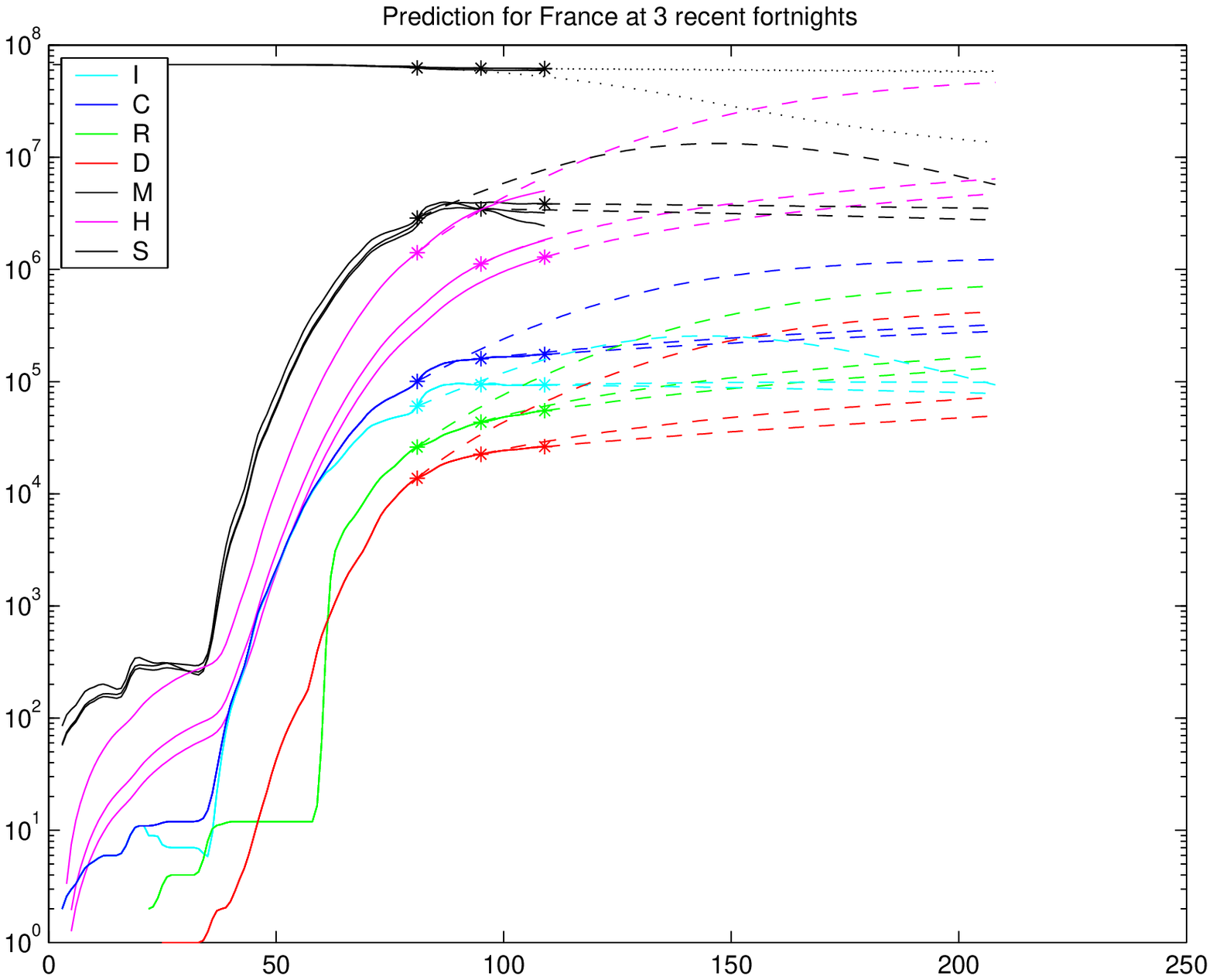}\\
  \includegraphics[width=12.0cm,height=5.0cm]{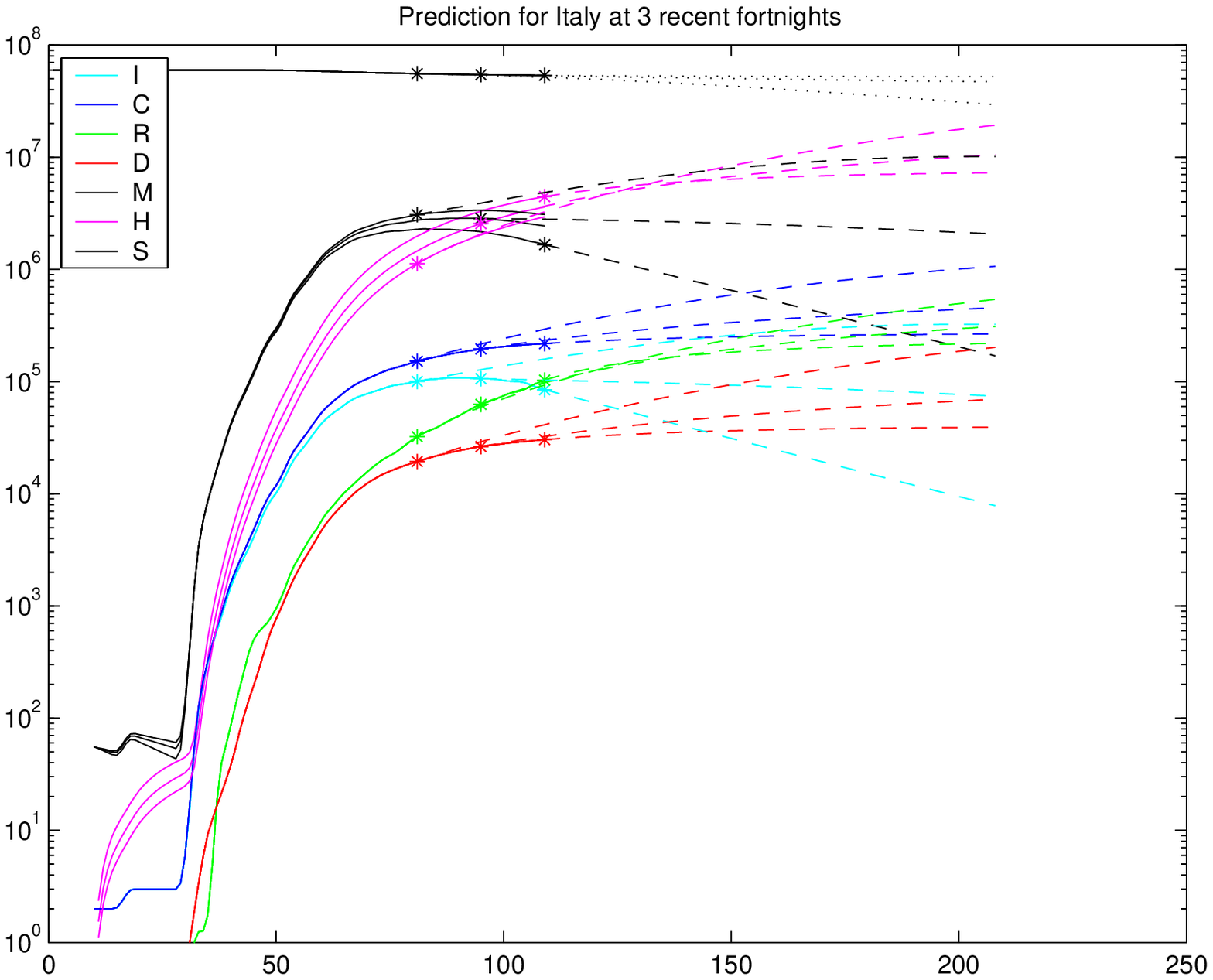}\\
  \includegraphics[width=12.0cm,height=5.0cm]{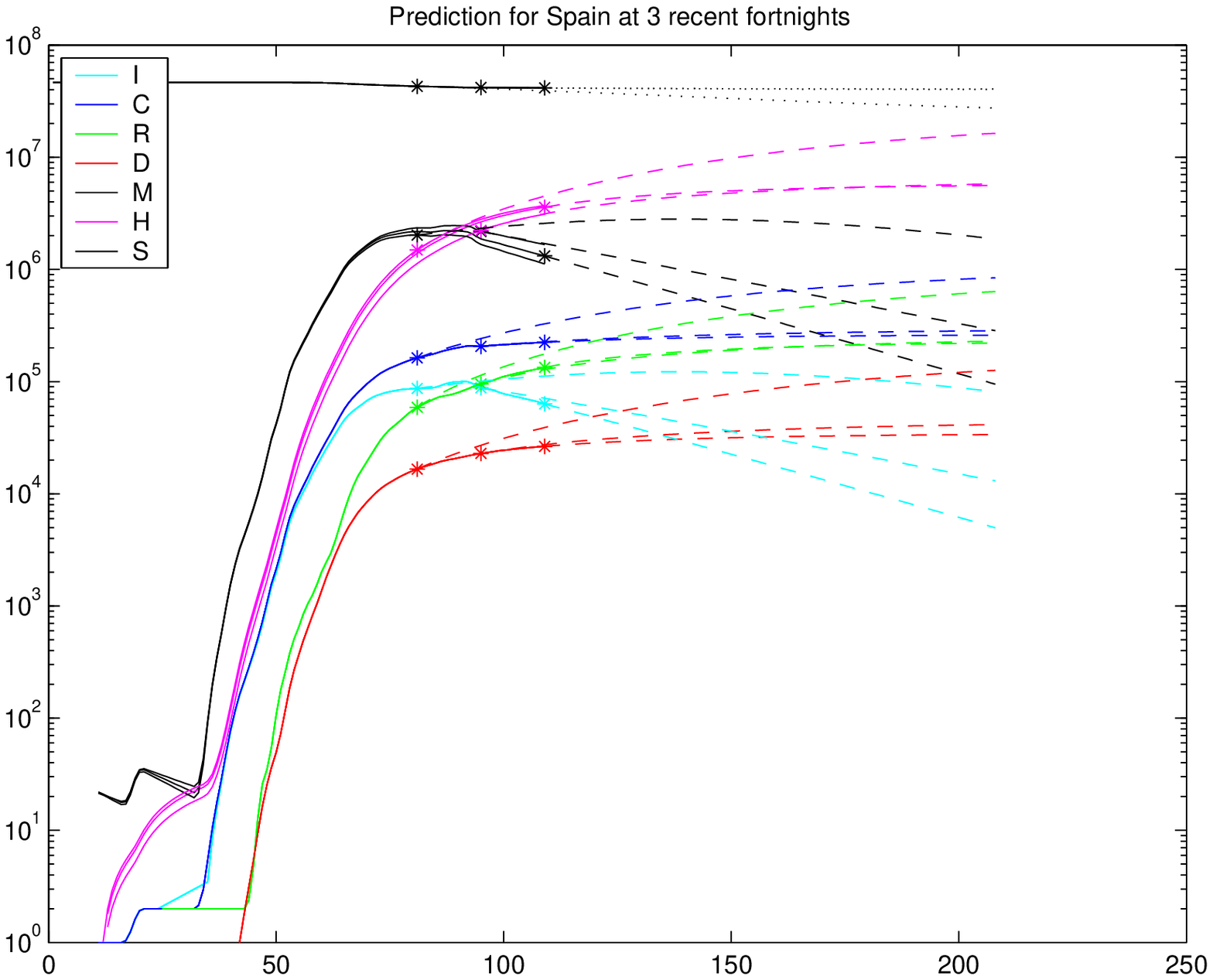}\\
  \includegraphics[width=12.0cm,height=5.0cm]{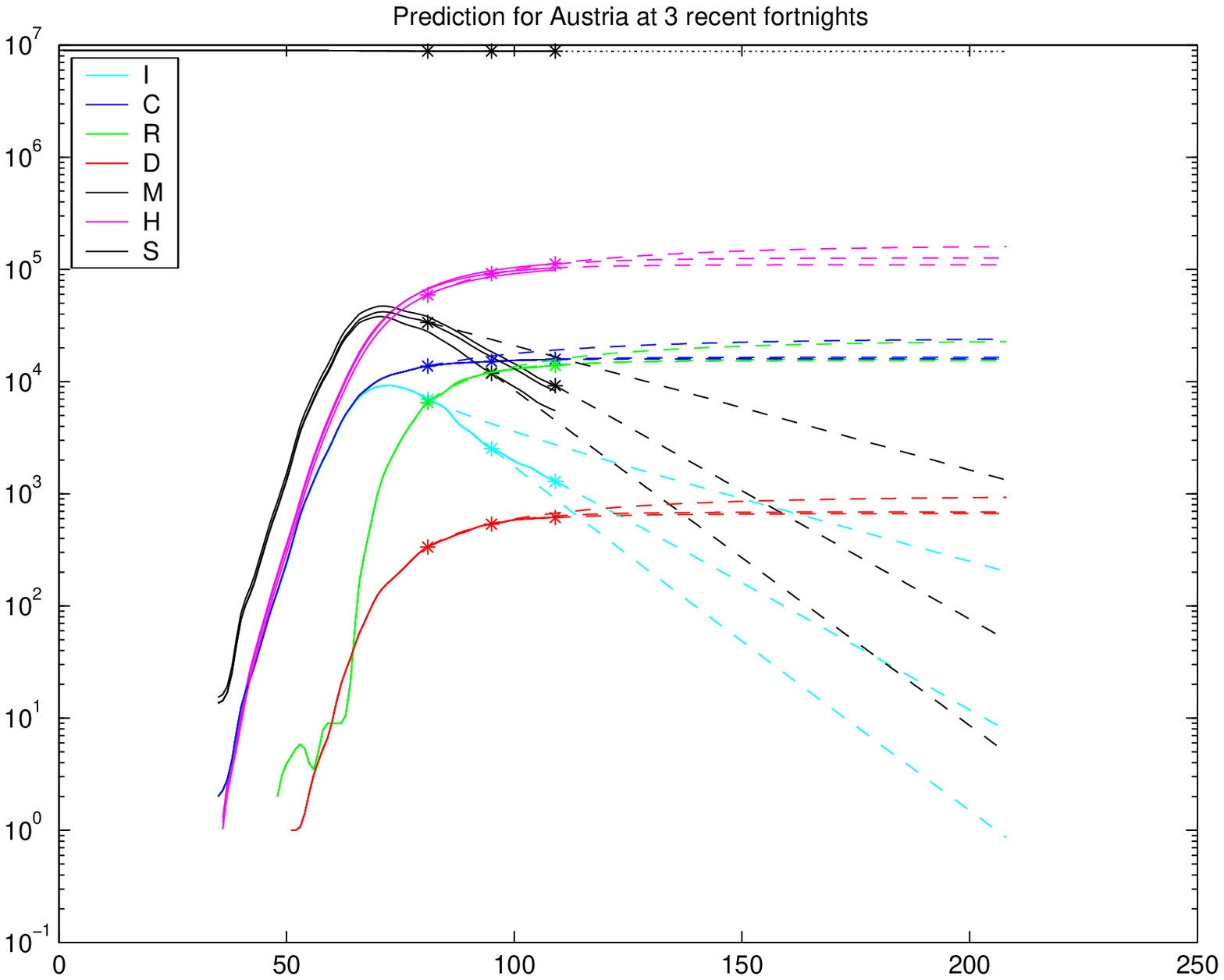}
  \caption{Predictions for countries France, Italy, Spain,
      and Austria on days
    109, 95, and 81
    \RSlabel{FigSMH109b3}} 
\end{center} 
\end{figure}
\biglf
All test runs were made for the infection fatality rate $\gamma_{IF}=0.005$,
the delay $k=21$ for estimating case fatalities, and a backlog of 7 days
when estimating constants out of recent values of time series.
The choice $\gamma_{IF}=0.005$ is somewhat between 
$0.56$\% from \RScite{anderheiden-buchholz:2020-1}, 
$0.66$\% from  \RScite{verity-et-al:2020-1}, and 
 $0.37$\% from \RScite{streeck:2020-1}. 
New information on infection fatality rates should be included as soon as they
are available.
\biglf
The Johns Hopkins data were smoothed by a double application
of the 1/4, 1/2, 1/4 filter on the logarithms,
like for Figure \RSref{FigR109}.
For UK
and Sweden, the data for the confirmed Recovered $R$ were replaced by
the $21$-day rule estimation via \RScite{schaback:2020-3}
and a backlog of $42$ days. The original data were too messy to be
useful, unfortunately. 
\biglf
For an early case in Germany, Figure \RSref{FigGermany67}
shows the prediction based on data of day 67, March 27 $^{th}$.
On the side, the figure  contains a wide variation of the 
starting value $H=N-S-C$
at the starting point, by multipliers between 1/32 and 32.
This has hardly any effect on the results. 
The peak of about 20 million
Infected is predicted on day 111, May 12$^{th}$,
with roughly a million Confirmed
and about 20000 casualties at the peak and
about 100.000 finally. Note that the real death count is about 7500
on May 11$^{th}$, and the prediction of the day, in Figure
\RSref{FigSMH109}, targets a final count of below 10000.
The parameter changes by political measures
turned out to be rather effective, like in many countries
that applied similar strategies. But since parts of the population
want to go back to their previous lifestyle, all of this is endangered,
and the figures should be monitored carefully.
\biglf
Of course, all of this
makes sense only under the assumption that reality follows the model,
in spite of all attempts to design a model that follows reality. 
\biglf
\begin{figure}
 \begin{center}
  \includegraphics[width=12.0cm,height=5.0cm]{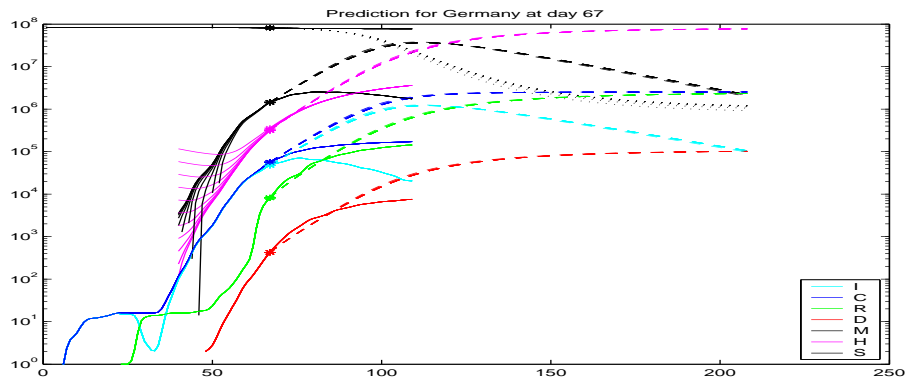}\\
  \caption{Prediction for Germany on day 67, March 27$^{th}$,
    with varying starting values. 
    \RSlabel{FigGermany67}} 
\end{center} 
\end{figure}

%****************************************************************
\section{Conclusion and Open Problems}\RSlabel{SecCaOP}
So far, the model presented here seems to be useful, combining
theory and practically available data. 
It is data-driven to a very large extent,
using only the infection fatality rate
from outside for prediction, and the approximation
\eref{eqggg} for calibration.
\biglf
On the downside, there is quite a number of shortcomings:
\begin{itemize}
\item Like the data themselves, the model needs regular updating.
As far as the Johns Hopkins data are concerned,
the model updates itself by using the latest data,
but it needs changes 
as soon as new
information on the hidden infections come in.
\item
  There may be better ways of estimating the hidden part of the
  epidemics. However, it will be easy to adapt the model to other
  pa\-ra\-me\-ter choices. If time series for the unknown variables
  get available, the model can easily be adapted to being data-driven
  by the new data.
\item The treatment of delays is unsatisfactory. In particular,
  infected persons get infected immediately, and the $k$-day rule is
  not followed at all places in the model.
  But the rule is violated as well in the data \RScite{schaback:2020-3}.
\item There is no stochastics involved, except for
  simple things like estimating constants by least squares,
  or for certain probabilistic arguments on the side, e.g.
  in section \RSref{SecEoCFR}. But it is not at all clear whether
  there are enough data to do a proper probabilistic analysis.
\item As long as there is no probabilistic analysis,
  there should be more simulations under perturbations of the data
  and the pa\-ra\-me\-ters. A few were included, e.g. for section
  \RSref{SecEoCFR} and Figures \RSref{FigcaseDRItaly109}
    and \RSref{FigGermany67},
but a large number was performed in the background when preparing the paper,
  making sure that results are stable, 
  but there are never too many test simulations. 
\item  Other models were not considered, e.g. the classical ones
  with delays
  \RScite{hoppenstaedt-waltman:1970-1,hoppenstaedt-waltman:1971-1}.
\item Under certain circumstances, epidemics
  do not show an exponential outbreak, in particular
if they hit only locally and a prepared population. See Figure
\RSref{FigGT} for the COVID-19 cases in Göttingen and vicinity.
\item Estimates for the peak time in section \RSref{SecBttP}
  need improvement.
\item Same for the underpinning of ``flattening the curve''
  in section \RSref{SecBtFtC}.
\end{itemize} 
MATLAB programs are available on request.
 \begin{figure}
 \begin{center}
  \includegraphics[width=8.0cm,height=10.0cm]{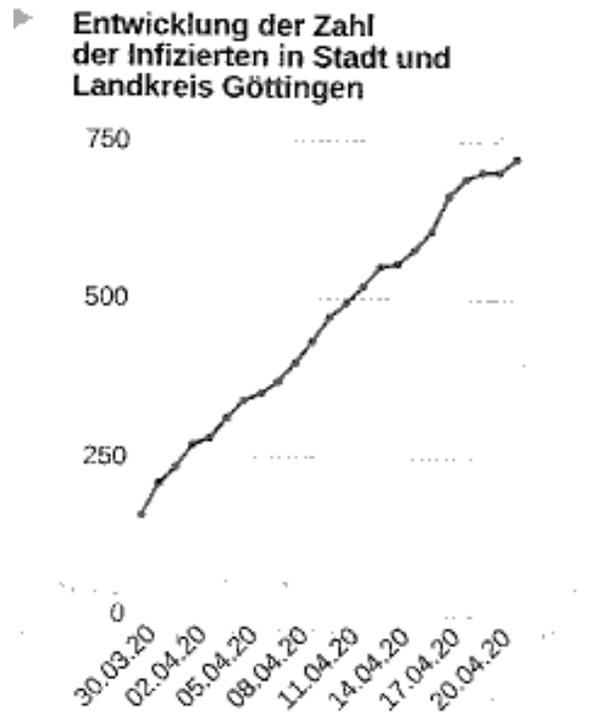}
  \caption{Infectious in Göttingen city and county,
    as of April 22nd, 2020 in the local newspaper ``Göttinger Tageblatt''.
    No exponential outbreak.
    \RSlabel{FigGT}}
\end{center} 
\end{figure}

%%%%%%%%%%%%%%%%%%%%%
%%%%%%%%%%%%%%%%%%%%%
%%%%%%%%%%%%%%%%%%%%%
%%%%%%%%%%%%%%%%%%%%%
%%%%%%%%%%%%%%%%%%%%%
%%%%%%%%%%%%%%%%%%%%%
%%%%%%%%%%%%%%%%%%%%%
%%%%%%%%%%%%%%%%%%%%%
\bibliographystyle{plain}

\begin{thebibliography}{10}
\bibitem{streeck:2020-1}
  Streeck et. al: Preliminary results from the Heinsberg outbreak,\\
  cited after G\"ottinger Tageblatt, May 5th, page 3.
\end{thebibliography}

\end{document}